\documentclass[11pt, a4paper]{article}
\pdfoutput=1
\usepackage{jheppub}

\usepackage{comment}

\usepackage{enumitem}

\usepackage{centernot} 
\usepackage{slashed} 

\usepackage[english]{babel}
\usepackage[utf8]{inputenc}  
\usepackage[T1]{fontenc}

\usepackage[normalem]{ulem}

\usepackage{subfig}

\usepackage{xcolor, graphicx}
\usepackage{tikz}
\usepackage{colortbl}

\usepackage{amsfonts, amssymb, amsthm} 
\DeclareFontFamily{U}{mathx}{\hyphenchar\font45}
\DeclareFontShape{U}{mathx}{m}{n}{
      <5> <6> <7> <8> <9> <10>
      <10.95> <12> <14.4> <17.28> <20.74> <24.88>
      mathx10
      }{}
\DeclareSymbolFont{mathx}{U}{mathx}{m}{n}
\DeclareFontSubstitution{U}{mathx}{m}{n}
\DeclareMathAccent{\widecheck}{0}{mathx}{"71}
\DeclareMathAccent{\wideparen}{0}{mathx}{"75}

\usepackage[tbtags]{mathtools}   
\numberwithin{equation}{section} 
\allowdisplaybreaks              

\usepackage{stmaryrd}
\usepackage{bbold}     

\makeatletter
\g@addto@macro\bfseries{\boldmath}
\makeatother

\usepackage{ifthen}
\newcommand{\dInt}[2][]{%
    \ifthenelse{\equal{#1}{}}
    {\ensuremath{\operatorname{d}{#2}\;}}
    {\ensuremath{\operatorname{d}^{#1}{#2}\;}}
}

\renewcommand{\imath}{i} 

\newcommand{\Proj}[1]{{\mathbb{P}_\text{#1}}}

\DeclareMathOperator{\Tr}{Tr}

\makeatletter
\renewcommand*\env@matrix[1][*\c@MaxMatrixCols c]{%
	\hskip -\arraycolsep
	\let\@ifnextchar\new@ifnextchar
	\array{#1}}
\makeatother

\usepackage[capitalise]{cleveref}

\let\oldFootnote\footnote
\newcommand\nextToken\relax
\renewcommand\footnote[1]{%
	\oldFootnote{#1}\futurelet\nextToken\isFootnote}
\newcommand\isFootnote{%
	\ifx\footnote\nextToken\textsuperscript{\normalfont,}\fi}

\usepackage[referable]{threeparttablex}

\usepackage{tabularx}
\newcolumntype{C}{>{\centering\arraybackslash}X}
\newcolumntype{R}{>{\raggedleft\arraybackslash}X}
\newcolumntype{L}{>{\raggedright\arraybackslash}X}

\usepackage{multirow, bigdelim}

\def\ChiQED/{$\chi$QED}

\newcommand{\be}{\begin{equation}}
\newcommand{\ee}{\end{equation}}
\newcommand{\bea}{\begin{eqnarray}}
\newcommand{\eea}{\end{eqnarray}}
\newcommand{\beas}{\begin{eqnarray*}}
\newcommand{\eeas}{\end{eqnarray*}}

\title{Two-loop application of the Breitenlohner-Maison/’t Hooft-Veltman scheme with non-anticommuting $\gamma_5$: Full renormalization and symmetry-restoring counterterms in an abelian chiral gauge theory}

\author[a]{Hermès Bélusca-Maïto,} 
\author[a]{Amon Ilakovac,} 
\author[b]{Paul Kühler,} 
\author[a,1]{Marija Mađor-Božinović\note{Corresponding author.},} 
\author[b]{and Dominik Stöckinger} 

\affiliation[a]{Department of Physics, University of Zagreb, Bijeni\v{c}ka cesta 32, HR-10000 Zagreb, Croatia}
\affiliation[b]{Institut für Kern- und Teilchenphysik, TU Dresden, Zellescher Weg 19, DE-01069 Dresden, Germany}

\emailAdd{hbelusca@phy.hr}
\emailAdd{ailakov@phy.hr}
\emailAdd{paul.kuehler@tu-dresden.de}
\emailAdd{mmadjor@phy.hr}
\emailAdd{Dominik.Stoeckinger@tu-dresden.de}

\abstract{
  We apply the BMHV scheme for non-anticommuting $\gamma_5$ to an abelian chiral gauge theory at the two-loop level. As our main result, we determine the full structure of symmetry-restoring counterterms up to the two-loop level. These counterterms turn out to have the same structure as at the one-loop level and a simple interpretation in terms of restoration of well-known Ward identities.
In addition, we show that the ultraviolet divergences cannot be canceled completely by counterterms generated
by field and parameter renormalization, and we determine needed UV divergent evanescent counterterms.  The paper establishes the two-loop methodology based on the quantum action principle and direct computations of Slavnov-Taylor identity breakings. The same method will be applicable to nonabelian gauge theories.
}

\begin{document}

\begin{flushright}
ZTF-EP-21-06 \\
\today
\end{flushright}

\maketitle
\flushbottom


\newpage

\section{Introduction}
\label{sect:Intro}

Dimensional Regularization (DReg)
\cite{Cicuta:1972jf,Bollini:1972ui,Ashmore:1972uj,tHooft:1972tcz}
is one of the most commonly employed schemes for practical calculations in perturbative quantum field theories.
However, in this scheme, the $\gamma_5$ Dirac matrix requires a special
treatment since not all its 4-dimensional properties have
straightforward $d$-dimensional extensions. This fact complicates
calculations, and various alternative treatments have been proposed; the issue of $\gamma_5$ has been known for a long time \cite{Chanowitz:1979zu,Kreimer:1989ke,Korner:1991sx,Kreimer:1993bh,Larin:1993tq,Trueman:1995ca,Chetyrkin:1997gb,Jegerlehner:2000dz} but is of increasing importance in current investigations, see Refs.\ \cite{Bednyakov:2015ooa,Zoller:2015tha,Bruque:2018bmy,Gnendiger:2017rfh,Poole:2019kcm,Poole:2019txl,Zerf:2019ynn,Ahmed:2020kme,Ahmed:2021spj,Cherchiglia:2020iug,TorresBobadilla:2020ekr,Cherchiglia:2021uce,Cherchiglia:2021yxz}
and Ref.\ \cite{Jegerlehner:2000dz} for an extensive overview of the
situation and further references.

In this paper, we follow the \emph{``Breitenlohner--Maison--'t~Hooft--Veltman''} (BMHV) scheme
\cite{tHooft:1972tcz, Akyeampong:1973xi,Akyeampong:1973vk,Akyeampong:1973vj}
of Dimensional Regularization. In this scheme $\gamma_5$ is non-anticommuting in $d$ dimensions, but the scheme is rigorously established at all orders
\cite{Breitenlohner:1975qe,Breitenlohner:1977hr,Breitenlohner:1975hg,Breitenlohner:1976te}.
Gauge invariance and the related BRST symmetry are broken in
intermediate steps by the modified algebraic relations.

For these reasons, the renormalization and counterterm structure in
the BMHV scheme involves several BMHV-specific complications: the
ultraviolet (UV) divergences cannot be cancelled by counterterms
generated by field and parameter renormalization; additional,
UV divergent evanescent counterterms (corresponding to operators which
vanish in strictly 4 dimensions) are needed; and the breaking of BRST
symmetry needs to be repaired by adding finite, symmetry-restoring
counterterms.

In a previous paper \cite{Belusca-Maito:2020ala} we have started a
research programme on the rigorous practical application of this BMHV
scheme to chiral gauge theories. In that reference we treated a
general non-abelian massless gauge theory with fermionic and scalar matter
fields, and we determined the full BMHV counterterm structure at the
one-loop level.
In particular we established a method
to
determine the required symmetry-restoring counterterms which
compensate the breaking of BRST symmetry.
The same method was developed and applied
earlier in
Ref.\ \cite{Martin:1999cc} to study one-loop symmetry breakings in the
BMHV scheme, and
Refs.\ \cite{Stockinger:2005gx,Hollik:2005nn,Stockinger:2018oxe} used
the same strategy to study supersymmetry in the context of dimensional
reduction up to the 3-loop level.
In the present work we present the first extension of this programme
to the two-loop level.
For the sake of clarity and to highlight conceptual and methodological
issues, we particularize our calculations to the case of an abelian
gauge theory with chiral fermions, a chiral QED (``\ChiQED/'') model.
Our goal is to determine the full two-loop structure of the special
counterterms in the BMHV scheme, i.e.\ the determine evanescent UV
divergences, the deviations from parameter and field renormalization,
and ultimately the symmetry-restoring counterterms.
The outline of the paper is as follows. After a brief reminder of
notation and properties of the scheme in \cref{sect:notation} we
define the abelian model in \cref{sect:model}. In this section we also
set up the Slavnov-Taylor identity corresponding to BRST invariance
and show that it is already broken at tree-level in the BMHV
scheme. \cref{sec:multiloopformulae} summarizes the general strategy
of renormalization and lays out the general procedure for finding
UV divergent and finite symmetry-restoring counterterms.
\cref{sect:Eval1LSCT,sect:BRSTrestoration}
contain the
one-loop counterterm results
for this model. Both the singular, including the evanescent ones, as
well as the BRST-restoring finite counterterms can also be derived by
particularizing our previously obtained generic results
\cite{Belusca-Maito:2020ala} to this model.
\cref{sect:Eval2LSCT} begins the two-loop analysis. It presents detailed results for the
UV divergences of subrenormalized two-loop Green functions, and
determines the required singular two-loop counterterms and their
relationship to field and parameter
renormalization. \cref{subsect:2LoopBRSTRestore} presents first the
evaluation of the two-loop breaking of the Slavnov-Taylor identity by
the regularization, using the method described in
\cref{sec:multiloopformulae} and
Ref.\ \cite{Belusca-Maito:2020ala}. It then presents the required
symmetry-restoring two-loop counterterms. We also provide a
consistency check by explicitly evaluating the analog of the usual QED
Ward identities for two-, three- and four-point functions and checking
that they are correctly restored as well.

\section{Generalities and Notations}
\label{sect:notation}

We re-employ the notations and conventions of \cite{Belusca-Maito:2020ala}.
The $d$-dimensional space is split into a direct sum of
$4$-dimensional and $d-4 = -2\epsilon$-dimensional subspaces. Lorentz
covariants are extended into this $d$-dimensional space and consist of
4-dimensional (denoted by bars: $\overline{\cdot}$\;) and
$(-2\epsilon)$-dimensional (also called ``evanescent'', denoted by
hats: $\widehat{\cdot}$\;) components.

This split is performed for any tensorial quantity; this includes the definition and properties of metric tensors%
\footnote{
	Our convention for the 4-dimensional metric signature is mostly minus, i.e. $(+1, -1, -1, -1)$.
}%
$g_{\mu\nu}$, $\bar{g}_{\mu\nu}$ and $\hat{g}_{\mu\nu}$,
of vectors $k_\mu$, $\bar{k}_\mu$ and $\hat{k}_\mu$,
and of the $\gamma$-matrices $\gamma^\mu$, $\bar{\gamma}^\mu$ and $\hat{\gamma}^\mu$.
In addition two intrinsically $4$-dimensional objects are defined and
appropriate properties for them are given. The first of them is the
Levi-Civita symbol $\epsilon_{\mu\nu\rho\sigma}$. The second one is
the $\gamma_5$ matrix. They are related by $\gamma_5 = \frac{-i}{4!}
\epsilon_{\mu\nu\rho\sigma} \gamma^\mu \gamma^\nu \gamma^\rho
\gamma^\sigma$.
In the BMHV scheme, the most important properties of the $\gamma_5$
matrix are its commutation and anticommutation relations with the
other $\gamma$-matrices, in particular
\begin{subequations}
\begin{align}
\label{eq:Gamma5DReg_A}
    \{\gamma_5, \bar{\gamma}^\mu\} &= 0 \, , \;&
    \{\gamma_5, \gamma^\mu\} &= \{\gamma_5, \hat{\gamma}^\mu\} = 2 \gamma_5 \hat{\gamma}^\mu \, , \;\\
    [\gamma_5, \hat{\gamma}^\mu] &= 0 \, , \;&
            [\gamma_5, \gamma^\mu] &= [\gamma_5, \bar{\gamma}^\mu] = 2 \gamma_5 \bar{\gamma}^\mu \, .
\end{align}
\end{subequations}
These relations will be used throughout all calculations in the present paper. They are
the root of the breaking of symmetries and the appearance of UV
divergences associated with purely evanescent operators.

\section{Right-Handed Chiral QED (\ChiQED/) and its Extension to $d$ Dimensions}
\label{sect:model}

The present paper is devoted to the first 2-loop application of the
method described in Ref.\ \cite{Belusca-Maito:2020ala}. We restrict
ourselves to the abelian $U(1)$ case without scalar fields and denote
the corresponding model as \ChiQED/. The model may be viewed either as
a chiral version of QED with purely right-handed fermion couplings, or
as a variant of the $U(1)$ part of the electroweak Standard Model.
Since the adjoint representation is trivial, trilinear and quartic
gauge boson interactions as well as ghost-gauge interaction are
absent. The fermionic generators are reduced to the hypercharge, with
opposite sign for the conjugate representation. The model is first
defined in 4 dimensions, then extended to $d$ dimensions, providing
the respective Lagrangian, BRST transformations and Slavnov-Taylor
identities, with motivation for particular choice for the evanescent
part of the fermion kinetic term and for the fermionic interaction
term. We then specify the BRST breaking of the model at
tree-level. Finally we collect the symmetry identities defining the
model at higher orders.

\subsection{\ChiQED/ in 4 Dimensions}

In \ChiQED/, the only generator is the hypercharge, which we can
assume to be diagonal,
$$\mathcal{Y}_{Rij}\equiv (\text{diag}\{ \mathcal{Y}_{R}^1, \dots, \mathcal{Y}_{R}^{N_f}\})_{ij},$$
where $N_f$ is the number of fermion flavours. The 4-dimensional classical Lagrangian of the model reads:
\begin{equation}
\label{eq:lagrangian}
\mathcal{L} =\imath \overline{\psi_R}_i \slashed{D}_{ij} {\psi_R}_j- \frac{1}{4} F^{\mu\nu} F_{\mu\nu} - \frac{1}{2 \xi} (\partial_\mu A^\mu)^2-\bar{c}\partial^2c+ \rho^\mu s{A_\mu} + \bar{R}^i s{{\psi_R}_i} + R^i s{\overline{\psi_R}_i},
\end{equation}
where only purely right-handed fermions $\psi_R$ appear. In general,
we use the standard right/left chirality projectors
$\Proj{R} = (\mathbb{1}+\gamma_5)/2$ and
$\Proj{L}=(\mathbb{1}-\gamma_5)/2$ and abbreviate
$\psi_{R/L}=\Proj{R/L}\psi$. The left-handed fermions $\psi_L$ are
thus decoupled from the theory.
The covariant derivative acting on the fermion field is defined in the
diagonal basis for couplings by
\begin{equation}
	D_{ij}^\mu = \partial^\mu \delta_{ij} - \imath e A^\mu {\mathcal{Y}_R}_{ij} \, ,
\end{equation}
and the field strength tensor is defined as
\begin{equation}
	F_{\mu\nu} = \partial_\mu A_\nu - \partial_\nu A_\mu \, .
\end{equation}
In order not to have anomalies in \ChiQED/, the following \emph{anomaly cancellation condition} is imposed on the hypercharge couplings,
\begin{equation}
\label{eq:AnomCond}
	\Tr(\mathcal{Y}_R^3) = 0 \, .
\end{equation}

Next, the Lagrangian contains an $R_\xi$ gauge fixing term with gauge
parameter $\xi$ and a corresponding
Faddeev-Popov ghost kinetic term.
The last three terms of \cref{eq:lagrangian} are the
BRST transformations of the physical fields, coupled to
external sources (or Batalin-Vilkovisky ``anti-fields'',
\cite{Batalin:1977pb,Batalin:1981jr,Batalin:1984jr}), where the
external sources do not transform under BRST transformations. The
non-vanishing BRST transformations are
\begin{subequations}
	\label{eq:BRST4}
	\begin{align}
	s{A_\mu} &= \partial_\mu c\, , \\
	s{\psi_i} &= s{{\psi_R}_i} = \imath \, e \, c \,{\mathcal{Y}_R}_{ij} {\psi_R}_j \, , \\    { }
	s{\overline{\psi}_i} &= s{\overline{\psi_R}_i} = \imath \, e \, \overline{\psi_R}_j c {\mathcal{Y}_R}_{ji} \, \\
	s{\overline{c}}&=B\equiv -\frac{1}{\xi}\partial A,\label{Bdefinition}
	\end{align}
\end{subequations}
where ``$s$'' is the generator of the BRST transformation, which acts
as a fermionic differential operator and is nilpotent for any linear
combination of fields. The last of these equations also introduces the
auxiliary Nakanishi-Lautrup field $B$, which is integrated
out from the action in \cref{eq:lagrangian} and in the rest of
this paper.
The 4-dimensional tree-level action
\begin{equation}
    S_0^{(4D)} = \int \dInt[4]{x} \mathcal{L}
\end{equation}
satisfies the following Slavnov-Taylor identity
\label{eq:STIS0}
\begin{equation}
    \mathcal{S}(S_0^{(4D)}) = 0 \, ,
\end{equation}
where the Slavnov-Taylor operation is given for a general functional $\mathcal{F}$ as
\begin{equation}\begin{split}
\label{eq:SofFDefinition}
	\mathcal{S}(\mathcal{F}) =
	\int \dInt[4]{x} \left(
	\frac{\delta \mathcal{F}}{\delta \rho^\mu} \frac{\delta \mathcal{F}}{\delta A_\mu} +
	\frac{\delta \mathcal{F}}{\delta \bar{R}^i} \frac{\delta \mathcal{F}}{\delta \psi_i} +
	\frac{\delta \mathcal{F}}{\delta R^i} \frac{\delta \mathcal{F}}{\delta \overline{\psi}_i} +
	B \frac{\delta \mathcal{F}}{\delta \bar{c}}
	\right)
	\, ,
\end{split}\end{equation}
where again $B$ is treated as an abbreviation to its value given in \cref{Bdefinition}.
As usual in the context of algebraic renormalization, several
additional functional identities hold. In particular
all functional derivatives of $S_0^{(4D)}$ with respect to the fields $c$,
$\bar{c}$ or $\rho^\mu$ are linear in the propagating
fields, and one may write down identities of the form $\delta
S_0^{(4D)}/\delta c(x)=\text{(linear expression)}.$ Such identities
may be required to hold at all orders as part of the definition of the
theory.\footnote{%
If $B$ is not integrated out, the same is true for the functional
derivative $\delta S_0^{(4D)}/\delta B(x)$.}
We highlight first the so-called ghost equation
\begin{align}
  \label{eq:GhostEq}
	 \bigg(\frac{\partial}{\partial\bar{c}} + \partial_\mu\frac{\partial}{\partial \rho_\mu}\bigg) S_0^{(4D)} &= 0 \, , 
\end{align}
which is a linear combination which has analogues also in the
non-abelian case.\footnote{%
It can be obtained in general from evaluating $\delta
  \mathcal{S}(S_0^{(4D)})/\delta B$ if the field $B$ is not
  eliminated.}
Second, the so-called antighost equation, based on $\delta
S_0^{(4D)}/\delta c(x)$, contains the essence of the original gauge
transformations. Combining it with the Slavnov-Taylor identity yields
the functional form of the abelian Ward
identity (for extensive discussions of the more general case and the
importance to the Standard Model and extensions see
e.g.\ \cite{Kraus:1997bi,Grassi:1999nb,Hollik:2002mv}). Here this
functional Ward
identity reads
\begin{align}
\label{TreelevelWI}
  \bigg(\partial^\mu\frac{\delta}{\delta A^\mu(x)}
  -ie {\mathcal{Y}}_{R}^{j}
  \sum_{\phi
  }(\pm)\phi(x)\frac{\delta}{\delta\phi(x)}
  \bigg)
  S_0^{(4D)}
  &=-\partial^2 B(x)
  \,.
\end{align}
The summation extends over the charged fermions and their sources,
$\phi\in\{{\psi_R}_j,\overline{\psi_R}_j,R^j,\bar{R}^j\}$, and the
signs are $+,-,+,-$, respectively.
Finally, we summarize in \cref{tbl:fields_quantum_numbers} a list of the quantum numbers (mass dimension, ghost number and (anti)commutativity) of the fields and the external sources of the theory, that are necessary for building the whole set of all possible renormalizable mass-dimension $\leq 4$ field-monomial operators with a given ghost number.

\begin{table}[h]
	\renewcommand*{\arraystretch}{1.2} 
	\centering
	\begin{tabular}{|r|*{7}{c}|cc|}
		\hline
		\multicolumn{1}{|r}{}
		                    & $A_\mu$ & $\overline{\psi}_i$, $\psi_i$ & $c$ & $\bar{c}$ & $B$ & $\rho^\mu$ & $R^i$, $\bar{R}^i$ & $\partial_\mu$ & $s$ \\
		\hline
		mass dimension      &  1      &  3/2                     &  0  &  2        &  2  &  3         &  5/2               &  1             &  0  \\
		ghost number        &  0      &  0                       &  1  & -1        &  0  & -1         & -1                 &  0             &  1  \\
		(anti)commutativity & +1      & -1                       & -1  & -1        & +1  & -1         & +1                 & +1             & -1  \\
		\hline
	\end{tabular}
	\caption{List of fields, external sources and operators, and their quantum numbers.}
	\label{tbl:fields_quantum_numbers}
\end{table}

\subsection{The \ChiQED/ in $d$ Dimensions and its BRST Breaking}

The extension of the \ChiQED/ model \cref{eq:lagrangian} to $d$ dimensions is not unique,
due to the fermionic kinetic and interaction terms.
Here we follow the procedure used in Ref.\ \cite{Belusca-Maito:2020ala}. The extension to $d$ dimensions requires fully $d$-dimensional fermion propagators, so as to ensure that Feynman diagrams involving fermions can be regularized. This is achieved by introducing a left-chiral $U(1)$-singlet fermion into the kinetic part of the Lagrangian, thus promoting the intrisically $4$-dimensional \ChiQED/ fermionic kinetic term to a full $d$-dimensional one.
On the other side the fermion-gauge boson interaction is chosen to be
fully chiral-projected, with right-handed fermions only. This
procedure, together with the straightforward extension of the other
terms in \cref{eq:lagrangian} to $d$ dimensions, leads to the
tree-level action $S_0$,
\begin{equation}
\label{eq:S0Def}
\begin{split}
	S_0 =\;& \int \dInt[d]{x} \bigg( \imath \overline{\psi}_i \slashed{\partial} {\psi}_i + e {\mathcal{Y}_R}_{ij} \overline{\psi_R}_i \slashed{A} {\psi_R}_j - \frac{1}{4} F^{\mu\nu} F_{\mu\nu}
	- \frac{1}{2 \xi} (\partial_\mu A^\mu)^2 \\
	&-\bar{c} \partial^2 c + \rho^\mu (\partial_\mu c) + \imath \, e \, \bar{R}^i c \,{\mathcal{Y}_R}_{ij} {\psi_R}_j + \imath \, e \, \overline{\psi_R}_i c {\mathcal{Y}_R}_{ij} R^j  \bigg)
	\, \\
	\equiv\;&
	\sum_i S^i_{\overline{\psi}\psi} + \sum_i \overline{S^i_{\overline{\psi}_R A \psi_R}}
	 + S_{AA}
	 + S_\text{g-fix} + S_{\bar{c} c}
	 + S_{\rho c}  + S_{\bar{R} c \psi_R} + S_{R c \overline{\psi_R}}
	\, ,
\end{split}
\end{equation}
where the last line introduces explicit abbreviations for each of the
eight terms in the action.

The rest of this subsection is devoted to the discussion of BRST
symmetry of the tree level action which follows the corresponding
discussion in Ref. \cite{Belusca-Maito:2020ala} for a generic
non-Abelian model. First we can define $d$-dimensional BRST
transformations and a $d$-dimensional Slavnov-Taylor operation
$\mathcal{S}_d$ by straightforward extensions of the $4$-dimensional
versions.  Then it is elementary to see that the $d$-dimensional
action may be written as the sum of two parts, an ``invariant'' and
an ``evanescent'' part,
\begin{subequations}
\begin{align}
    S_0 &= S_{0,\text{inv}} + S_{0,\text{evan}} \, , \\
    S_{0,\text{evan}} &= \int \dInt[d]{x} \imath
           \overline{\psi}_i \widehat{\slashed{\partial}} {\psi}_i \, .
\end{align}
\end{subequations}
The first part is BRST invariant even in $d$ dimensions, i.e.\ it
satisfies
\begin{subequations}
  \begin{align}
  s_d S_{0,\text{inv}}&=0\,,\\
  \mathcal{S}_d(S_{0,\text{inv}}) &=0\,.
\end{align}
\end{subequations}
The second part $S_{0,\text{evan}}$ consists solely of one single,
evanescent fermion kinetic term, but it breaks $d$-dimensional BRST
invariance and the tree-level Slavnov-Taylor identity,
%
\begin{subequations}
  \begin{align}
	s_d S_0 &= s_d S_{0,\text{evan}} \equiv \widehat{\Delta} \, , \\
	S_d (S_0) &= \widehat{\Delta} \, .
  \end{align}
\end{subequations}
The breaking $\widehat{\Delta}$ is an integrated evanescent operator, comprised of one ghost and two fermions,
\begin{equation}\begin{split}
\label{eq:BRSTTreeBreaking}
	\widehat{\Delta}
	&= \int \dInt[d]{x} e\, {\mathcal{Y}_R}_{ij} \, c \, \left\{
		\overline{\psi}_i \left(\overset{\leftarrow}{\widehat{\slashed{\partial}}} \Proj{R} + \overset{\rightarrow}{\widehat{\slashed{\partial}}} \Proj{L}\right) \psi_j
		\right\}
	\equiv \int \dInt[d]{x} \widehat{\Delta}(x)
	\, ,
\end{split}\end{equation}
and generates an interaction vertex whose Feynman rule (with all momenta incoming) is
\begin{equation}
\begin{tabular}{rl}
	\raisebox{-40pt}{\includegraphics[scale=0.6]{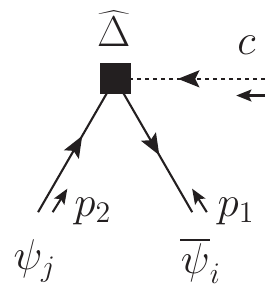}} &
	$\begin{aligned}
		&= e \, {\mathcal{Y}_R}_{ij} \left(\widehat{\slashed{p_1}} \Proj{R} + \widehat{\slashed{p_2}} \Proj{L} \right)
		\, .
	\end{aligned}$
\end{tabular}
\end{equation}

For later use we also need the expression for the linearized Slavnov-Taylor operator \cite{Piguet:1995er}, $b_d$, defined such that
\begin{align}
  \mathcal{S}_d(S_0 + \hbar \mathcal{F}) &= \mathcal{S}_d(S_0) + \hbar \, b_d{\mathcal{F}} + \mathcal{O}(\hbar^2)\,.
\end{align}
Its functional definition in \ChiQED/ and its relation to $s_d$ are:
\begin{equation}\begin{split}
\label{eq:bdDefinition}
	b_d &=
	\int \dInt[d]{x} \left(
	\frac{\delta S_0}{\delta \rho^\mu} \frac{\delta}{\delta A_\mu}
	+ \frac{\delta S_0}{\delta A_\mu} \frac{\delta}{\delta \rho^\mu}
	+ \frac{\delta S_0}{\delta \bar{R}^i} \frac{\delta}{\delta \psi_i} + \frac{\delta S_0}{\delta \psi_i} \frac{\delta}{\delta \bar{R}^i}
	+ \frac{\delta S_0}{\delta R^i} \frac{\delta}{\delta \overline{\psi}_i} + \frac{\delta S_0}{\delta \overline{\psi}_i} \frac{\delta}{\delta R^i}
	+ B \frac{\delta}{\delta \bar{c}}
	\right)
	\\
	&= s_d +
	\int \dInt[d]{x} \left(
	\frac{\delta S_0}{\delta A_\mu} \frac{\delta}{\delta \rho^\mu}
	+ \frac{\delta S_0}{\delta \psi_i} \frac{\delta}{\delta \bar{R}^i}
	+ \frac{\delta S_0}{\delta \overline{\psi}_i} \frac{\delta}{\delta R^i} \right)
	\, .
\end{split}\end{equation}

\subsection{Defining Symmetry Relations for the Renormalized Theory}
\label{sect:definingsymmetries}
At higher orders a set of symmetry identities can be imposed on the
finite, renormalized theory. These identities may be viewed as part of
the definition of the model; they constrain the
regularization/renormalization procedure and particularly determine
the symmetry-restoring counterterms. Here we collect the relevant
symmetry identities which are the basis of the subsequent sections. All
following identities are valid at tree level
by construction. In principle, it is crucial to establish that they can be
fulfilled also at higher orders. For the present model this is clear
from the general analysis of algebraic renormalization of gauge
theories\footnote{%
  See Refs.\ \cite{Becchi:1974md,Piguet:1980nr,Kraus:1995jk,Haussling:1996rq} for important treatments of abelian
theories in such contexts and Refs.\ \cite{Piguet:1980nr,Piguet:1995er} for
general overviews.}
and the anomaly condition \cref{eq:AnomCond}.

All identities are formulated as functional identities for the fully
renormalized, finite and 4-dimensional
effective action $\Gamma_\text{ren}$, which
formally satisfies
$\Gamma_\text{ren} = S_0^{(4D)} + \mathcal{O}(\hbar)$.
The first and most important symmetry is BRST invariance, which is
expressed as the Slavnov-Taylor identity
\begin{equation}
\label{GeneralSTI}
\mathcal{S}(\Gamma_\text{ren}) = 0
\end{equation}
for the renormalized theory. In addition, we require a set of more
trivial relations
\begin{align}
\label{TrivialIdentities}
\frac{\delta\Gamma_\text{ren}}{\delta c(x)}
&= \frac{\delta S_0^{(4D)}}{\delta c(x)} \, ,
&
\frac{\delta\Gamma_\text{ren}}{\delta\bar c(x)}
&= \frac{\delta S_0^{(4D)}}{\delta \bar c(x)} \, ,
&
\frac{\delta\Gamma_\text{ren}}{\delta \rho^\mu(x)}
&= \frac{\delta S_0^{(4D)}}{\delta \rho^\mu(x)} \, .
\end{align}
These identities correspond
to  the absence of higher-order corrections involving the fields $c$,
$\bar{c}$, $\rho^\mu$ (a similar identity for the $B$-field is valid
in case $B$ is not yet eliminated).
They can be imposed since the respective derivatives of the tree-level
action
are linear in the dynamical fields as described between
\cref{eq:SofFDefinition,TreelevelWI}.

Like at tree level, the Ward identity
\begin{equation}\label{eq:GeneralWI}
\bigg(\partial^\mu\frac{\delta}{\delta A^\mu(x)}-ie {\mathcal{Y}}_{R}^{j}\sum_{\phi}(\pm)\phi(x)\frac{\delta}{\delta\phi(x)}\bigg)\Gamma_\text{ren} = -\partial^2 B(x) \, ,
\end{equation}
is an automatic consequence of the Slavnov-Taylor identity \cref{GeneralSTI}
combined with the antighost equation in \cref{TrivialIdentities}.
It is not
    manifestly valid at
    higher orders but it will be automatically valid once the
    Slavnov-Taylor identity holds. In fact we will see that the
    breaking and restoration of the Slavnov-Taylor identity can be
    well interpreted in terms of the Ward identity.

In what follows we will only refer to BRST
  invariance and the Slavnov-Taylor identity, which are the most
  important symmetry requirements. The requirements \cref{TrivialIdentities} are
  manifestly
  valid at all steps and individually for the regularized Green
  functions and for the counterterms.
\section{Multiloop Regularization and Renormalization Formulae}
\label{sec:multiloopformulae}
Dimensional regularization using the BMHV scheme inevitably breaks BRST symmetry, which therefore has to be restored at any order in perturbation
for ensuring the consistency of the theory. In this section we collect
the general formulae governing the construction of the renormalized
theory and the procedure for finding singular (i.e.\ UV divergent) and
finite symmetry-restoring counterterms.
The calculational details at the one-loop and two-loop level will be
given in the following sections.

In general, counterterms contain UV divergent (``singular'') and
finite contributions.
As noted in our previous paper \cite{Belusca-Maito:2020ala},
in dimensional regularization it is useful to further subdivide the
counterterms into five types:
singular BRST invariant and noninvariant (evanescent or
non-evanescent) counterterms, finite  BRST
invariant and noninvariant (BRST restoring) counterterms as well as
finite evanescent counterterms,
\begin{equation}
S_\text{ct} = S_\text{sct,inv}+ S_\text{sct,noninv}+ S_\text{fct,inv}+
S_\text{fct,restore}+S_\text{fct,evan}\equiv S_\text{sct}+S_\text{fct} \, .
\label{eq:Sctdec}
\end{equation}
In this and the following equations, symbols with no index
denote all-order quantities. For the following perturbative expressions
we will also use an upper
index $i$ for quantities of precisely order $i$, an upper index $(i)$
for quantities up to and including order $i$. For example, the
counterterm action and the bare action can be split as
\begin{align}
  S_\text{bare}&=S_0+S_\text{ct},&
  S_\text{ct}&=\sum_{i=1}^\infty S_\text{ct}^i,&
  S_\text{ct}^{(i)}&=\sum_{j=1}^iS_\text{ct}^j\,.
\end{align}
The perturbative construction of the effective action in dimensional
regularization and renormalization is performed iteratively at each
order of $\hbar$ (or loops), starting from the tree-level action $S_0$
of order $\hbar^0$. Then, at each higher loop order
$i\ge1$ a counterterm action $S_\text{ct}^{i}$ has to be constructed. The
counterterms are subject to the two conditions that the renormalized
theory is UV finite and in agreement with all required
symmetries listed in \cref{sect:definingsymmetries}.\footnote{%
  As mentioned in  \cref{sect:definingsymmetries}, in what follows
we will only refer to BRST
  invariance and the Slavnov-Taylor identity, which are the most
  important symmetry requirements. The other
related symmetry requirements \cref{TrivialIdentities} are manifestly
valid at all steps.  }
In general,  at each order $i$ one may distinguish Green functions at
various levels of regularization, partial or full renormalization. Of
particular importance are ``subloop-renormalized'' Green functions and
the corresponding effective action. To keep the notation simple in the
present paper we use the symbol $\Gamma^i$ for this
subloop-renormalized effective action of order $i$. By
definition this is obtained at order $i$ by using Feynman rules from
the tree-level action and counterterms up to order $i-1$. By
constructing and including singular counterterms of the order $i$ we
obtain the quantity
\begin{align}
 \Gamma^i +  S_\text{sct}^i &= \text{finite for }\epsilon\to0\,.
\label{Ssctdefinition}
\end{align}
This equation determines the singular counterterms unambiguously,
including their evanescent parts, introduced in \cref{eq:Sctdec}.
By also including additional, finite counterterms of the order $i$ we
obtain
\begin{align}
  \Gamma_\text{DReg}^{i} := \Gamma^{i} +  S_\text{sct}^i +
  S_\text{fct}^i\,.
\end{align}
This resulting effective action is finite at this order and
essentially renormalized but still contains the variable $\epsilon$
and evanescent quantities. The fully renormalized effective action is
given by taking the limit $d\to4$ and by setting all evanescent
quantities to zero. This operation is denoted as
\begin{align}
  \Gamma_\text{ren}^{i} := \text{LIM}_{d\to 4} \Gamma_\text{DReg}^{i}\,.
\end{align}

The basic procedure to determine the finite
counterterms $S_\text{fct}^i$, specifically their symmetry-restoring
part, is as follows. The ultimate symmetry requirement is the
Slavnov-Taylor identity expressing BRST invariance for the fully
renormalized theory, which can
be written as
\begin{equation}
\label{eq:SdGamD4d}
	\text{LIM}_{d\to 4} \big(\mathcal{S}_d (\Gamma_\text{DReg})\big) = 0 \, .
\end{equation}
As discussed in detail in Ref.\ \cite{Belusca-Maito:2020ala} there are
several possibilities to extract the symmetry-restoring counterterms
from this equation. Like in that reference, we choose again to use the regularized quantum
action principle \cite{Breitenlohner:1977hr}, which allows to
rewrite\footnote{%
   The same equation has been presented specifically for the one-loop case
   in Ref.\ \cite{Belusca-Maito:2020ala} and for the general case in
   Ref.\ \cite{Stockinger:2005gx}. Ref.\ \cite{Martin:1999cc}
   presents a slightly different version. All versions of the equation
   become equal in the present context of an abelian gauge theory
   where there are no counterterms involving external fields.}
\begin{equation}\begin{split}
\label{eq:SdGamosb}
	\mathcal{S}_d (\Gamma_\text{DReg})
		&= (\widehat{\Delta} + \Delta_\text{ct}) \cdot
        \Gamma_\text{DReg}
	\, ,
\end{split}\end{equation}
where the insertions $\widehat{\Delta}$  and $\Delta_\text{ct}$ in the
present abelian theory are given as
\begin{subequations}
  \label{eq:Deltas}
\begin{align}
	\widehat{\Delta} & = \mathcal{S}_d( S_0) =s_d S_0 \, ,
	\label{eq:DelpertE0}\\
	\widehat{\Delta}+	\Delta_\text{ct} & =\mathcal{S}_d( S_0+S_\text{ct}) \, ,
	\label{eq:DelpertEtotal}\\
	\Delta_\text{ct} &\equiv s_d S_\text{ct} \,.
\label{eq:DelpertE}
\end{align}
\end{subequations}
The first two equations are valid in general, the third one is valid
in the present context because, as we will see in the concrete
calculations, there will be no counterterms involving external fields.
The previous equations can be plugged into \cref{eq:SdGamD4d} and perturbatively expanded
at the order $i$. This leads to
\begin{align}
\label{eq:SdGamExp4d}
	\text{LIM}_{d\to 4}\,\Big(\widehat{\Delta}\cdot\Gamma_\text{DReg}^i
	+ \sum_{k=1}^{i-1} \Delta_\text{ct}^k\cdot\Gamma_\text{DReg}^{i-k} + \Delta_\text{ct}^i\Big) = 0 \, ,
	\qquad \text{for}\; i \geq 1
	\, ,
\end{align}
which explicitly exhibits the genuine $i$-loop counterterm via
$\Delta_\text{ct}^i$ (see Eq. \eqref{eq:Deltas}). The fact that the limit $d\to4$ exists provides
a consistency check on the divergent part of $\Delta_\text{ct}^i$, which contains
the singular counterterms $S_\text{sct}^i$. The finite part of the
equation determines the finite part of $\Delta_\text{ct}^i$. This
determines the desired finite counterterms not unambiguously. Rather,
their symmetry-restoring parts are fixed, while it remains possible to
add finite symmetric counterterms (to adjust renormalization
conditions) and finite evanescent counterterms, which are not needed
in the following.

\section{Evaluation of the One-Loop Singular Counterterm Action $S_\text{sct}^{(1)}$ in \ChiQED/ versus QED}
\label{sect:Eval1LSCT}

We start by evaluating the one-loop (order $\hbar^1$) singular
counterterm action $S_\text{sct}^{(1)}$, defined from the divergent
parts of the one-loop diagrams constructed with the Feynman rules of
the tree-level action $S_0$. These counterterms are basically
determined by \cref{Ssctdefinition} at the one-loop level, and
they  will be part of the dimensionally-regularized one-loop effective
action $\Gamma_\text{DReg}$.

The calculations are performed in $d = 4 - 2\epsilon$ dimensions.
We use notational conventions from \cite{Belusca-Maito:2020ala}.
Here and in the rest of the paper, the necessary Feynman diagrams have been computed using the \verb|Mathematica| packages \verb|FeynArts|~\cite{Hahn:2000kx} and \verb|FeynCalc|~\cite{Mertig:1990an,Shtabovenko:2016sxi}; the $\epsilon$-expansion of the amplitudes has been cross-checked using the \verb|FeynCalc|'s interface \verb|FeynHelpers|~\cite{Shtabovenko:2016whf} to \verb|Package-X|~\cite{Patel:2016fam}.

Since intermediate results can be obtained from the presentation of
Ref.\ \cite{Belusca-Maito:2020ala}, we immediately provide the full result for the
singular one-loop counterterm action. It reads
\begin{align}
\label{eq:SingularCT1Loop}
	\begin{split}
	S_\text{sct,\ChiQED/}^{1} =\;&
		\frac{-\hbar \, e^2}{16 \pi^2 \epsilon} \left(
		\frac{2 \Tr(\mathcal{Y}_R^2)}{3} \overline{S_{AA}}
		+\xi \, \sum_j (\mathcal{Y}_R^j)^2 \left( \overline{S^j_{\overline{\psi}\psi_R}} + \overline{S^j_{\overline{\psi_R} A \psi_R}} \right)
		\right.\\
		&\left. + \frac{\Tr(\mathcal{Y}_R^2)}{3} \int \dInt[d]{x} \frac{1}{2} \bar{A}_\mu \widehat{\partial}^2 \bar{A}^\mu \right)
	\, ,
	\end{split}
\end{align}
and it may be compared to the corresponding result of ordinary QED
with Dirac fermions of charges $\mathcal{Y}$,
\begin{align}
  \label{eq:SingularCT1LoopQED}
  S_\text{sct,QED}^{1} =\;&
		\frac{-\hbar \, e^2}{16 \pi^2 \epsilon} \left(
		\frac{4 \Tr(\mathcal{Y}^2)}{3} S_{AA}
		+ \xi \, \sum_j (\mathcal{Y}^j)^2 \left( S^j_{\overline{\psi}\psi}
		+ S^j_{\overline{\psi} A \psi} \right) \right)
	\,.
\end{align}
Notice that we restore explicit $\hbar$ order for every final result of the counterterm action from now on.
Most of the monomials have already been introduced; the bar in $ \overline{S_{AA}}
$ designates the fully 4-dimensional version of $S_{AA}$, and the
additional terms $\overline{S^i_{\overline{\psi}\psi_R}}$,
$\overline{S^i_{\overline{\psi_R} A \psi_R}}$ are the fully
right-chiral-projected equivalents to their usual $d$-dimensional
versions,
\begin{subequations}
\begin{align}
	& \overline{S^i_{\overline{\psi}\psi_R}} =
		\int \dInt[d]{x} \imath \overline{\psi}_i \overline{\slashed{\partial}} \Proj{R} \psi_i
		\equiv  \int \dInt[d]{x} \frac{\imath}{2} \overline{\psi}_i \overset{\leftrightarrow}{\overline{\slashed{\partial}}} \Proj{R} \psi_i \, ,
	\\
	& \overline{S^i_{\overline{\psi_R} A \psi_R}} =
		\int \dInt[d]{x} e\mathcal{Y}_R^i \overline{\psi}_i \overline{\slashed{A}} \Proj{R} \psi_i \, .
\end{align}
\end{subequations}
In both models there are three kinds of UV divergent Green functions,
corresponding to the photon self energy, the fermion self energy and the
fermion--photon interaction.
The results \cref{eq:SingularCT1Loop,eq:SingularCT1LoopQED} differ in three characteristic ways. Clearly, in \ChiQED/
there are half as many fermionic degrees of freedom, hence the fermion
loop contributions to the photon self energy generate the prefactor
$2/3$ instead of $4/3$. In addition, the purely right-handed nature of
the interaction leads to a purely evanescent divergent non-transverse
contribution to the photon self energy in the second line of
\cref{eq:SingularCT1Loop}. Finally,
the fermion self energy and the fermion-photon interaction receive
only  purely 4-dimensional right-handed corrections in \ChiQED/, while
in (non-chiral) QED these contributions remain $d$-dimensional.

Both \ChiQED/ and ordinary QED models are abelian, and as a result
there are no loop corrections involving ghosts or external BRST source
fields. This property reflects the identities
\cref{TrivialIdentities} and persists at all orders.
This implies in particular that the linearized Slavnov-Taylor operator
$b_d$ reduces to the BRST operator $s_d$ when acting on the loop
contributions of the effective action.

As in Ref.\  \cite{Belusca-Maito:2020ala} we can re-express the result
for the singular one-loop counterterms $S_\text{sct}^{1}$ in a structure
reminiscent of
the one appearing in the usual renormalization transformations, where
fields renormalize multiplicatively as $\varphi \to \sqrt{Z_\varphi}\,
\varphi$, $Z_\varphi \equiv 1 + \delta Z_\varphi$, and the coupling
constant renormalizes additively as $e \to e + \delta e$.
The sum of the singular counterterms can be written as
\begin{equation}
\label{eq:Ssct1L}
	S_\text{sct}^{1} = S_\text{sct,inv}^{1} + S_\text{sct,evan}^{1} \, ,
\end{equation}
where the first term arises in the usual way from a renormalization
transformation, while the second term has a different structure. In
detail, the first term can be obtained by applying the renormalization
transformation
$S_{0,\text{inv}} \longrightarrow S_{0,\text{inv}} + S_\text{ct,inv}$,
and it is given by
\begin{equation}
	S_\text{ct,inv}^{1} =
	\frac{\delta Z_A^{1}}{2} L_A +
	\frac{\delta Z_c^{1}}{2} L_c +
	\frac{\delta Z_{\psi_{Rj}}^{1}}{2} \overline{L_{\psi_{Rj}}} +
	\frac{\delta e^{1}}{e} L_{e} \, .
\end{equation}
The one-loop renormalization constants $\delta Z_\varphi$, $\delta e$
agree with the usual ones
(see e.g. \cite{Machacek:1983tz,Machacek:1983fi,Machacek:1984zw}) and read
\begin{subequations}
\begin{align}
\label{eq:delZ1LoopConst}
	\delta Z_A^1 &= \delta Z_c^1 = -2 \frac{\delta e^1}{e} \, ,
        \\
        \label{eq:RenConst1LoopFirst}
\delta Z_A^{1} &=
	\frac{-\hbar \, e^2}{16 \pi^2 \epsilon} \frac{2 \Tr(\mathcal{Y}_R^2)}{3}
\, ,
\\
\delta Z_{\psi_{R\,j}}^{1} &=
	\frac{-\hbar \, e^2}{16 \pi^2 \epsilon} \xi (\mathcal{Y}_R^j)^2
\, .\label{eq:RenConst1LoopLast}
\end{align}
\end{subequations}
The first of these relations again reflects
\cref{TrivialIdentities} as in ordinary QED.
The $L_\varphi$ functionals corresponding to field renormalizations
can be written in various ways, either as a field-numbering operators
acting on the tree-level action or as total $b_d$-variations or in terms of
the monomials of \cref{eq:S0Def}. Here we provide the results in the
form
\begin{subequations}
\begin{align}
    L_A &= b_d \int \dInt[d]{x} \widetilde{\rho}^\mu A_\mu
         = 2 S_{AA} + \overline{S_{\overline{\psi} A \psi_R}} - S_{\bar{c} c} - S_{\rho c}
        \, ,
    \\
	\intertext{where $\widetilde{\rho}^\mu = \rho^\mu +
          \partial^\mu \bar{c}$ is the natural combination arising
          from the ghost equation (\ref{eq:GhostEq});}
    L_c &= 
    \int \dInt[d]{x}c(x)\frac{\delta S_0}{\delta c(x)}
         = S_{\bar{c} c} + S_{\rho c} + S_{\bar{R} c \psi_R} + S_{R c \overline{\psi_R}}
        \, ,\label{ugly}
    \\
    \begin{split}
    L_{\psi_R} &= - b_d \int \dInt[d]{x} (\bar{R}^i \Proj{R} \psi_i + \overline{\psi}_i \Proj{L} R^i) \\
        &=
        2 \left( \int \dInt[d]{x} \imath \overline{\psi}_i \overline{\slashed{\partial}} \Proj{R} \psi_i + \overline{S_{\overline{\psi} A \psi_R}} \right) + \imath \overline{\psi}_i \widehat{\slashed{\partial}} \psi_i \equiv \overline{L_{\psi_R}} + S_{0,\text{evan}}=\sum_i L_{\psi_{Ri}}
        \, .
    \end{split}
\end{align}
\end{subequations}
The $L_{e}$ functional corresponding to renormalization of the
physical coupling can be expressed in terms of the monomials of
\cref{eq:S0Def} or related to the field renormalization functionals as
\begin{equation}
    L_{e} = e \frac{\partial S_0}{\partial e} = \overline{S_{\overline{\psi} A \psi_R}} + S_{\bar{R} c \psi_R} + S_{R c \overline{\psi_R}} = L_c + L_A - 2 S_{AA}
    \, .
\end{equation}

Despite the non-nilpotency of $b_d$, several of the $L_\varphi$ are actually $b_d$-invariant in the following sense:
\begin{align}
  b_d L_A &= 0 \, , &
  b_d \overline{L_{\psi_R}} &= 0 \, .
\end{align}
In contrast, $L_c$ is not $b_d$-invariant in this sense%
\footnote{
	This fact appears to be in contradiction with a claim made in \cite{Martin:1999cc}.
};
instead, it is easy to see that
\begin{align}
	b_d L_c &= \widehat{\Delta} \, ,
\end{align}
with the same breaking as in \cref{eq:BRSTTreeBreaking}. As a result, also $L_{e}$, corresponding to gauge coupling renormalization, is not $b_d$-invariant.
Note, however, that in the limit $d \to 4$ and evanescent terms vanishing, all the $L_\varphi$ functionals presented here become invariant under the linear $b$ transformation in 4 dimensions.

Finally, the evanescent counterterms appearing in \cref{eq:Ssct1L} can be written as
\begin{equation}
\label{eq:Ssctevan}
\begin{split}
    S_\text{sct,evan}^{1} =\;&
        \frac{-\hbar \, e^2}{16 \pi^2 \epsilon} \frac{\Tr(\mathcal{Y}_R^2)}{3} \left( 2 (\overline{S_{AA}} - S_{AA}) + \int \dInt[d]{x} \frac{1}{2} \bar{A}^\mu \widehat{\partial}^2 \bar{A}_\mu \right)
    \, .
\end{split}\end{equation}
For later use we record the corresponding BRST breaking of the
singular one-loop counterterms. This breaking originates solely from
the evanescent non-invariant second term of
$S_\text{sct,evan}^{1}$ and is given by
\begin{equation}
\label{eq:bdSsct1L}
\begin{split}
\Delta_\text{sct}^1 = s_d  S_\text{sct}^{1}
	&= -\frac{\hbar}{16 \pi^2 \epsilon }\frac{e^2 \Tr(\mathcal{Y}_R^2)}{3}  \, \int \dInt[d]{x} \,(\overline{\partial}_\mu c) \, (\widehat{\partial}^2 \bar{A}^\mu)
	\, .
\end{split}
\end{equation}

\section{BRST Symmetry Breaking and its Restoration; Evaluation of the One-Loop Finite Counterterm Action $S_\text{fct}^{1}$}
\label{sect:BRSTrestoration}

In the previous section we determined the singular counterterms
action $S_\text{sct}^1$, \cref{eq:Ssct1L}. Here we discuss the determination of
symmetry-restoring counterterms $S_\text{fct}^1$ at the one-loop level. We follow the
general procedure outlined in \cref{sec:multiloopformulae} but
will be brief since
the computation is essentially a special case of the one presented in
Ref.\ \cite{Belusca-Maito:2020ala}. We begin by specializing the
general formulae of \cref{sec:multiloopformulae} to the
one-loop case, then we present the results and a brief discussion.
At the one-loop level the structure of renormalization can be written as
\begin{subequations}
\begin{align}
 \Gamma_\text{DReg}^{(1)} &= \Gamma^{(1)} + S_\text{sct}^1 + S_\text{fct}^1 \, ,
\label{eq:GammaDR1_2}
\\
	\Delta_\text{ct}^1 &= \mathcal{S}_d (S_0+S_\text{ct})^1 \, , 
\end{align}
\end{subequations}
where the counterterms are subject to the conditions discussed in
\cref{sec:multiloopformulae}, which here simplify to
\begin{subequations}
  \begin{align}
S^1_\text{sct} +\Gamma^1_\text{div} &=0\, ,
\label{eq:S1sct_1L}
\\
 \big(\widehat{\Delta}\cdot\Gamma^1_\text{} +
 \Delta_\text{ct}^1\big)_\text{div} &= 0 \, ,
\label{eq:S1sct_1Lcheck}
\\
\text{LIM}_{d\to 4} \big(\widehat{\Delta}\cdot\Gamma^1_\text{} +
 \Delta_\text{ct}^1\big)_\text{fin} &= 0 \, .
\label{eq:S1fct_1Lv1}
\end{align}
\end{subequations}
Here the subscripts $\text{`div,fin'}$ refer to the pure $1/\epsilon$
pole part and the $\epsilon$-independent part, respectively. Compared
to \cref{eq:SdGamExp4d} we dropped the index $\text{`DReg'}$ because the
one-loop insertions arise from genuine one-loop diagrams and not from
one-loop counterterms. Equation
(\ref{eq:S1sct_1L}) has already been satisfied in the previous section, and
\cref{eq:S1sct_1Lcheck} must automatically hold by construction, providing a
consistency check. The last
equation determines the finite symmetry-restoring counterterms, with a
remaining ambiguity of changing finite symmetric or evanescent
counterterms.
The equation can also be written as
\begin{align}
N \Big[\widehat{\Delta}\cdot\Gamma^1_\text{}\Big] + \Delta_\text{fct}^1 &= 0 \, ,
\label{eq:S1fct_1L}
\end{align}
which implicitly fixes the choice of the finite, evanescent
counterterms.
This version of the equation uses the result (\ref{eq:bdSsct1L}) that  the
BRST variation of the one-loop singular counterterms contains no
finite term (which could in principle arise from the evaluation of
$s_d$), hence $\Delta_\text{ct}|_\text{fin} = \Delta_\text{fct}$.
The symbol $N[{\cal O}]$ denotes the Zimmermann-like definition
\cite{Zimmermann:1972te,Zimmermann:1972tv,Lowenstein:1971vf,Piguet:1980nr} of a
renormalized local operator (also called \emph{``normal product''}),
defined as an insertion of a local operator ${\cal O}$ and followed,
in the context of Dimensional Regularization and Renormalization, by a
minimal subtraction prescription \cite{Collins:1974da}.

\begin{figure}
\begin{minipage}{\textwidth}
	\centering
	\includegraphics[scale=0.5]{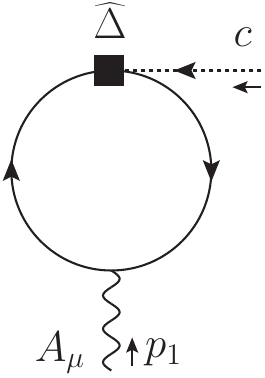}
	\hfill
	\raisebox{+5pt}{\includegraphics[scale=0.5]{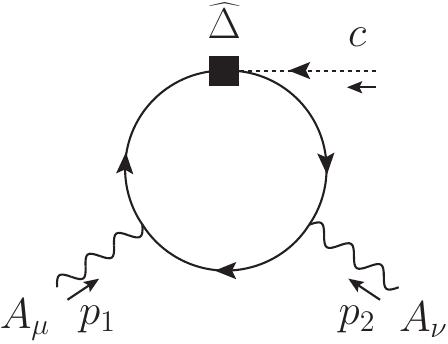}}
	\hfill
	\raisebox{-10pt}{\includegraphics[scale=0.5]{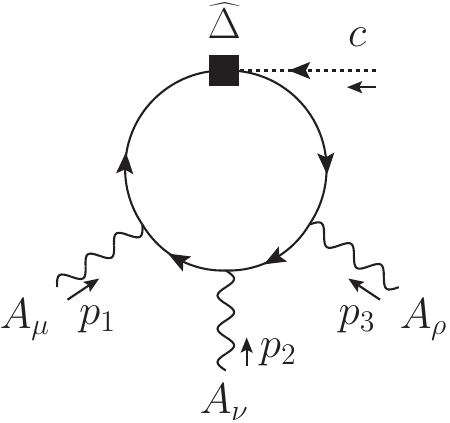}}
	\hfill
	\includegraphics[scale=0.5]{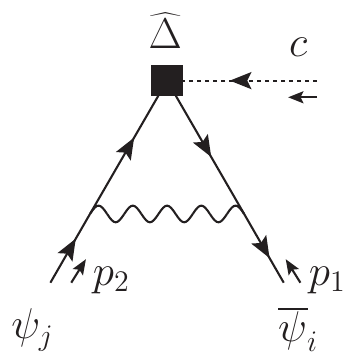}
\end{minipage}
\caption{\label{fig:oneloopinsertions} The four one-loop diagrams contributing to
  $\widehat{\Delta}\cdot\Gamma^1_\text{}$ in a relevant way. Only the first
diagram provides  UV divergent contributions, which are
evanescent. All diagrams provide UV finite non-evanescent
contributions, i.e.\ contributions which remain in the $\text{LIM}_{d\to 4}$.}
\end{figure}

In order to determine the finite counterterms we need to  compute the
quantity $\widehat{\Delta}\cdot\Gamma^1_\text{}$, corresponding to the
breaking of the Slavnov-Taylor identity or BRST symmetry by one-loop
regularized Green functions. This is
given by one-loop Feynman diagrams with one insertion of the vertex
$\widehat{\Delta}$, the BRST breaking of the d-dimensional action given in
\cref{eq:Deltas}. In principle, infinitely many Feynman diagrams
can give a nonzero result. However in most cases the result is purely
evanescent or of  order $\epsilon$. Only power-counting divergent
diagrams can lead to a result which contributes to the above
equations, i.e. which contains either a $1/\epsilon$ pole or which is
finite and survives in the $\text{LIM}_{d\to 4}$. The four
contributing diagrams are shown in \cref{fig:oneloopinsertions}.

The result of all these diagrams can be compactly written as an insertion in the
effective action in terms of field monomials of ghost number one, as
\begin{equation}
\label{eq:Delfct1L} 
\begin{split}
\widehat{\Delta}\cdot\Gamma^1 =\;&
    \frac{\hbar}{16\pi^2} \int \dInt[d]{x} \bigg[
       \, \frac{e^2 \Tr(\mathcal{Y}_R^2)}{3}\left( \frac{1}{\epsilon}(\overline{\partial}_\mu c) \, (\widehat{\partial}^2 \bar{A}^\mu)
	 + (\overline{\partial}_\mu c)  (\overline{\partial}^2 \bar{A}^\mu) \right) \\
    &+ \frac{e^4 \Tr(\mathcal{Y}_R^4)}{3} \, c \, \overline{\partial}_\mu  (\bar{A}^\mu \bar{A}^2)
		- \frac{5+\xi}{6} e^3 \sum_{j}(\mathcal{Y}_R^j)^3  \, c \, \overline{\partial}^\mu (\overline{\psi}_j \overline{\gamma}_\mu \Proj{R} \psi_j)\bigg]
    \, .
\end{split}
\end{equation}
In this equation, further terms of order $\epsilon$ and evanescent
terms of order $\epsilon^0$ have been omitted.
The first two terms correspond to the first diagram of
\cref{fig:oneloopinsertions}; they involve an evanescent UV
divergence and a UV finite,
non-evanescent term.
Their interpretation is the violation of the Slavnov-Taylor identity for the photon self energy (describing essentially its transversality).
The other terms are UV finite and
non-evanescent. They correspond in an obvious way to the third and fourth diagrams of \cref{fig:oneloopinsertions}, and they correspond to the violation of the Slavnov-Taylor identities involving the photon 4-point function and the fermion--photon interaction, respectively.
The contribution coming from the second diagram, $\big(\widehat{\Delta} \cdot \Gamma\big)_{AAc}^{1}$, turns out to be proportional to $\Tr(\mathcal{Y}_R^3) \epsilon^{\mu\nu\rho\sigma}$, i.e. proportional to the anomaly coefficient and the Levi-Civita symbol.
It therefore forms an essential anomaly and could not be canceled by any
counterterm. Here, it is zero in view of our assumption
\cref{eq:AnomCond}.

We can first use this result to carry out the check of the divergent
contributions \cref{eq:S1sct_1Lcheck}. The singular counterterms given in
\cref{eq:Ssct1L} are BRST invariant up to one evanescent contribution exhibited in
\cref{eq:Ssctevan}, and the resulting BRST breaking
$\Delta_\text{sct}^1$ is given in \cref{eq:bdSsct1L}. Hence, we
see that indeed \cref{eq:S1sct_1Lcheck} is fulfilled, as it
should be.

Now we turn to the determination of the finite counterterms
$S^{(1)}_\text{fct}$, which serve to restore the Slavnov-Taylor
identity in the form of Eqs.\ (\ref{eq:S1fct_1Lv1}) or
(\ref{eq:S1fct_1L}).
Explicit inspection of the breaking (\ref{eq:Delfct1L}) reveals that
the following ansatz is correct:
\begin{equation}
\label{eq:Sfct1L}
\begin{split}
	S^1_\text{fct} =
	\frac{\hbar}{16\pi^2} \int \dInt[4]{x} &\Bigg\{
	\frac{-e^2 \Tr(\mathcal{Y}_R^2)}{6} \bar{A} \cdot (\overline{\partial}^2 \bar{A})
	+ \frac{e^4 \Tr(\mathcal{Y}_R^4)}{12} (\bar{A}^2)^2
	\\
    & + \frac{5+\xi}{6} e^2 \sum_j (\mathcal{Y}_R^j)^2 \imath \overline{\psi}_j \overline{\gamma}^\mu \overline{\partial}_\mu \Proj{R} \psi_j \Bigg\}
	\, .
\end{split}
\end{equation}
Each of the three terms has a clear interpretation. The first restores
transversality of the photon self energy, the second restores the Ward
identity relation for the quartic photon interaction. The third term
restores the Ward identity between the fermion self energy and its
photon interaction.

The choice of the symmetry-restoring counterterms is constructed such that it satisfies
\begin{align}
  s_d S_\text{fct}^1 &=-N[\widehat{\Delta}]\cdot\Gamma^1\,,
\end{align}
where the right-hand side corresponds to the purely finite,
non-evanescent part of \cref{eq:Delfct1L}. Further, the counterterms do not
depend on external source fields, which implies the identities
\begin{align}
  s_d S_\text{fct}^1 &= b_d S_\text{fct}^1 = \mathcal{S}_d
  (S_0+S_\text{fct})^1 = \Delta_\text{fct}^1\,.
\end{align}
Hence the counterterms $S_\text{fct}^1$ restore the symmetry, and all equations
(\ref{eq:S1fct_1Lv1}, \ref{eq:S1fct_1L}) and ultimately (\ref{eq:SdGamD4d})
are valid at the one-loop level. As an additional byproduct the
simplification announced in \cref{eq:DelpertE} is established
at this order.
As mentioned above, the finite counterterms are not uniquely
fixed. One can add any BRST-symmetric term to these finite
counterterms without spoiling the restoration of the BRST
symmetry. This is required to fulfil specific renormalization
conditions but is not further pursued in the present paper.

Further, the finite counterterms are defined as purely
four-dimensional quantities. This corresponds to our
requirement (\ref{eq:S1fct_1L}). As discussed there, one may change
the finite counterterms by evanescent contributions which vanish in
the $\text{LIM}_{d\to 4}$. This means that it would be allowed e.g.\ to
change the counterterms by extending them to $d$ dimensions, i.e.\ to
replace some or all of the $\bar{A}_\mu$ and $\bar{\partial}_\mu$ by
full $A_\mu$ and $\partial_\mu$. Such changes are irrelevant for pure
one-loop discussions, however once the counterterms are inserted into
higher-loop diagrams the changes matter and might change the form of
two-loop results, corresponding to different renormalization schemes.


As stated around \cref{eq:S1fct_1L}, in this work we stick to
the choice of keeping the finite counterterms in
their four-dimensional form when being used in 2-loop calculations, as
vertex insertions.

\section{Evaluation of the Two-Loop Singular Counterterm Action $S_\text{sct}^{2}$ in \ChiQED/ versus QED}
\label{sect:Eval2LSCT}

In this section we determine the
UV divergences of the subrenormalized two-loop (order $\hbar^2$) Green
functions. According to \cref{Ssctdefinition} these define the
singular two-loop counterterm action $S_\text{sct}^{2}$.  The
calculations are performed in
$d = 4 - 2\epsilon$ dimensions, and in the Feynman gauge $\xi=1$, and
the results are
compared with the corresponding results for ordinary QED.

\subsection{A Comment on the Calculation Procedure}

The calculational procedure uses the same tools already mentioned in
\cref{sect:Eval1LSCT}. In addition, we compute two-loop self energy
integrals using \verb|TARCER| \cite{Mertig:1998vk}. Divergences of
three-point functions are obtained using two different approaches. In
the first approach we reduce the expressions effectively to self energies by
setting one external momentum to zero and proceed with \verb|TARCER|. This
is justified since we are interested in the UV divergences which are
known to be local and independent of external momenta for the diagrams of our interest (after
subrenormalization). This approach fails in case zero external momenta
induce infrared divergences. In this case, we use a
UV/IR-decomposition \cite{Misiak:1994zw,Chetyrkin:1997fm,Luthe:2017ttg} where effectively all external momenta vanish and propagators
become massive. The resulting integrals become massive self energies without
external momenta, i.e.\ massive vacuum integrals. Whenever different
approaches can be applied we use both, and the results agree.


\subsection{List of Divergent Two-Loop Green Functions}

Like at the one-loop level there are three kinds of UV divergent Green functions,
corresponding to the photon self energy, the fermion self energy and the
fermion--photon interaction. Here we first present the explicit
results for each subrenormalized two-loop Green function separately
and both for \ChiQED/ and ordinary QED.
The blobs shown in the diagrams represent the sum of the all possible
subrenormalized two-loop corrections, i.e.\ two-loop diagrams with
tree-level vertices and one-loop diagrams with singular and  finite
BRST-restoring counterterm insertions.

\noindent
\textbf{Gauge boson self energy:}\; 
\raisebox{-20pt}{\includegraphics[scale=0.6]{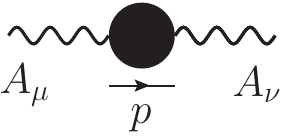}}
\begin{subequations}
  \begin{align}
\label{eq:QGBSE}
	\imath \widetilde{\Gamma}_{AA}^{\nu \mu}(p)|^{2}_\text{div, \ChiQED/} &=
		\frac{\imath e^4}{256 \pi^4} \frac{\Tr(\mathcal{Y}_R^4)}{3} \left[ \frac{2}{\epsilon} (\overline{p}^\mu \overline{p}^\nu - \overline{p}^2 \overline{g}^{\mu\nu}) + \left( \frac{17}{24 \epsilon} - \frac{1}{2 \epsilon^2} \right) \widehat{p}^2 \overline{g}^{\mu\nu} \right]
	\, ,
	\\
	\imath \widetilde{\Gamma}_{AA}^{\nu \mu}(p)|^{2}_\text{div, QED} &=
		\frac{\imath e^4}{256 \pi^4 \epsilon} 2 \, \Tr(\mathcal{Y}^4) (p^\mu p^\nu - p^2 g^{\mu\nu})
	\, .
  \end{align}
\end{subequations}

\noindent
\textbf{Fermion self energy:}\; 
\raisebox{-20pt}{\includegraphics[scale=0.6]{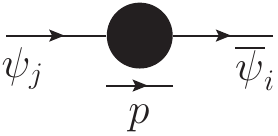}}
\begin{subequations}
\begin{align}
	\imath \widetilde{\Gamma}_{\psi \overline{\psi}}^{ji}(p)|^{2}_\text{div, \ChiQED/} &=
		\frac{-\imath e^4}{256 \pi^4} \left[ \frac{(\mathcal{Y}_R^2)^{ij}\,\Tr(\mathcal{Y}_R^2)}{9 \epsilon} + (\mathcal{Y}_R^4)^{ij} \left( \frac{7}{12 \epsilon} + \frac{1}{2 \epsilon^2} \right) \right] \overline{\centernot{p}}\; \Proj{R}
	\, ,
	\\
	\imath \widetilde{\Gamma}_{\psi \overline{\psi}}^{ji}(p)|^{2}_\text{div, QED} &=
		\frac{-\imath e^4}{256 \pi^4} \left[ \frac{(\mathcal{Y}^2)^{ij}\,\Tr(\mathcal{Y}^2)}{\epsilon} + (\mathcal{Y}^4)^{ij} \left( \frac{3}{4 \epsilon} + \frac{1}{2 \epsilon^2} \right) \right] \centernot{p}
	\, .
\end{align}
\end{subequations}


\noindent
\textbf{Fermion-gauge boson interaction:}\;
\raisebox{-25pt}{\includegraphics[scale=0.6]{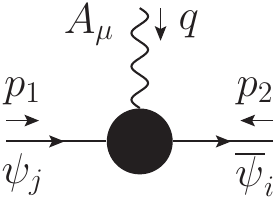}}
\begin{subequations}
\begin{align}
	\imath \widetilde{\Gamma}_{\psi \overline{\psi} A}^{ji,\mu}|^{2}_\text{div, \ChiQED/} &=
		\frac{-\imath e^5}{256 \pi^4} \left[ \frac{(\mathcal{Y}_R^2)^{ij}\,\Tr(\mathcal{Y}_R^3)}{\epsilon}
		- \frac{(\mathcal{Y}_R^3)^{ij}\,\Tr(\mathcal{Y}_R^2)}{9 \epsilon} + (\mathcal{Y}_R^5)^{ij} \left( \frac{17}{12 \epsilon}
		+ \frac{1}{2 \epsilon^2} \right) \right] \overline{\gamma}^\mu\, \Proj{R}
	\, ,
	\\
	\imath \widetilde{\Gamma}_{\psi \overline{\psi} A}^{ji,\mu}|^{2}_\text{div, QED} &=
		\frac{-\imath e^5}{256 \pi^4} \left[ \frac{(\mathcal{Y}^3)^{ij}\,\Tr(\mathcal{Y}^2)}{\epsilon} + (\mathcal{Y}^5)^{ij} \left( \frac{3}{4 \epsilon}
		+ \frac{1}{2 \epsilon^2} \right) \right] \gamma^\mu
	\, .
\end{align}
\end{subequations}
The first term with $\Tr(\mathcal{Y}_R^3)=0$ does not contribute due to the previously imposed anomaly cancellation condition.
\\
\\
\noindent
\textbf{Three- and four-photon interactions:}
The triple-photon interaction amplitude is equal to zero for QED models, while it is finite and purely evanescent for \ChiQED/. The four-photon interaction amplitude is finite and does not provide any singular counterterm.

\subsection{Singular Two-Loop Counterterms}

From the singular part of the two-loop diagrams listed above
we obtain the singular counterterm action at the two-loop level, which
cancels these divergences,
\begin{equation}
\label{eq:SingularCT2Loop}
\begin{split}
	S^2_\text{sct} =\;&
	-\Bigg(\frac{\hbar \, e^2}{16 \pi^2}\Bigg)^2 \frac{\Tr(\mathcal{Y}_R^4)}{3} \left[ \frac{2}{\epsilon} \overline{S_{AA}} + \left( \frac{1}{4\epsilon^2} - \frac{17}{48\epsilon} \right) \int \dInt[d]{x} \bar{A}_\mu \widehat{\partial}^2\bar{A}^\mu \right]
	\\
	& +\Bigg(\frac{\hbar \, e^2}{16 \pi^2}\Bigg)^2  \sum_{j}(\mathcal{Y}_R^j)^2
		\left[ \left(\frac{1}{2\epsilon^2} + \frac{17}{12\epsilon} \right) (\mathcal{Y}_R^j)^2
		- \frac{1}{9\epsilon} \Tr(\mathcal{Y}_R^2) \right]
		\left( \overline{S^j_{\overline{\psi}\psi_R}} + \overline{S^j_{\overline{\psi_R} A \psi_R}} \right)
	\\
	& - \Bigg(\frac{\hbar \, e^2}{16 \pi^2}\Bigg)^2  \sum_{j}\frac{(\mathcal{Y}_R^j)^2}{3\epsilon} \left( \frac{5}{2}(\mathcal{Y}_R^j)^2 - \frac{2}{3} \Tr(\mathcal{Y}_R^2) \right) \overline{S^j_{\overline{\psi}\psi_R}}
	\ .
\end{split}
\end{equation}
Its structure is the same as at the one-loop level, see
\cref{eq:SingularCT1Loop}, corresponding to counterterms to the
three divergent kinds of Green functions. Again purely 4-dimensional
terms appear, as well as an evanescent contribution to the photon self energy.
A conceptually new feature compared to the one-loop case is
the term in the last line of
$S_\text{sct}^2$, which  breaks BRST
invariance by a non-evanescent amount.

In the following we again re-express the result using renormalization
transformations. Because of the last term it is not possible
to split the two-loop singular counterterms into a BRST invariant plus
an evanescent part.  We can write
\begin{equation}
\label{eq:Ssct2L}
	S_\text{sct}^2 = S_\text{sct,inv}^2 + S_\text{sct,break}^2 \, .
\end{equation}
The first term is BRST invariant and arises from the renormalization transformation
$S_{0,\text{inv}} \longrightarrow S_{0,\text{inv}} + S_\text{ct,inv}$,
and is given by
\begin{equation}\begin{split}
\label{eq:Sctinvstructure2Loop}
	S_\text{sct,inv}^2 =\;&
	\frac{\delta Z_A^2}{2} L_A +
	\frac{\delta Z_c^2}{2} L_c +
	\frac{\delta Z_{\psi_{Rj}}^2}{2} \overline{L_{\psi_{Rj}}} +
	\frac{\delta e^2}{e} L_{e}
\end{split}\end{equation}
with two-loop renormalization constants $\delta Z_\varphi^2$, $\delta
e^2$.\footnote{%
  Note that the superscript in the notation $\delta e^2$ refers to the 2-loop
  contribution to $\delta e$, not to a square!}
The split (\ref{eq:Ssct2L}) is not unique, but we reasonably choose to
keep the ``trivial'' identities (\ref{TrivialIdentities}) valid for both
$S_\text{sct,inv}^2 $ and $S_\text{sct,break}^2$ individually
(which simply means the counterterms do not contain the ghost or
source fields) and
to allow only two-point functions in $S_\text{sct,break}^2$.
Then the renormalization constants are given by
\begin{subequations}
\begin{align}
\label{eq:delZ2LoopConst}
	\delta Z_A^2 &= \delta Z_c^2 = -2 \frac{\delta e^2}{e} \, ,
        \\
        \label{eq:RenConst2LoopFirst}
\delta Z_A^2 &=
	-\frac{e^4}{256 \pi^4 \epsilon} \frac{2 \Tr(\mathcal{Y}_R^4)}{3}
\, ,
\\
\delta Z_{\psi_{Rj}}^2 &=
	\frac{e^4}{256 \pi^4} (\mathcal{Y}_R^j)^2
		\left[ \left(\frac{1}{2\epsilon^2} + \frac{17}{12\epsilon} \right) (\mathcal{Y}_R^j)^2
		- \frac{1}{9\epsilon} \Tr(\mathcal{Y}_R^2) \right]
\, .
\end{align}
\end{subequations}
Like at the one-loop level the first equation (\ref{eq:delZ2LoopConst}) here reflects the
validity of the trivial identities
equation (\ref{TrivialIdentities}) and the analog of the usual QED Ward
identity on the level of $S_\text{sct,inv}^2$.
The results for the other renormalization constants differ from the
ones in the literature obtained without the BMHV scheme, see
e.g.\ Refs.\  \cite{Machacek:1983tz,Machacek:1983fi,Machacek:1984zw}. This
difference reflects the modified relationship between the
renormalization-group $\beta$ functions and singular counterterms in
the BMHV scheme, see the discussions in
Refs.\ \cite{Belusca-Maito:2020ala,Schubert:1989wg}. A detailed investigation of this issue
will be presented in a forthcoming publication.

The BRST-breaking singular counterterms appearing in \cref{eq:Ssct2L} can be written as
\begin{equation}
\label{eq:SsctevanTL}
\begin{split}
	S_\text{sct,break}^2 =\;&
		-\Bigg(\frac{\hbar \, e^2}{16 \pi^2}\Bigg)^2  \frac{1}{\epsilon} \frac{\Tr(\mathcal{Y}_R^4)}{3} \left( 2 (\overline{S_{AA}} - S_{AA})
		+ \left( \frac{1}{2\epsilon} - \frac{17}{24} \right)
		\int \dInt[d]{x} \frac{1}{2} \bar{A}^\mu \widehat{\partial}^2 \bar{A}_\mu \right)
		\\
		& - \Bigg(\frac{\hbar \, e^2}{16 \pi^2}\Bigg)^2 \frac{1}{3\epsilon} \sum_{j}(\mathcal{Y}_R^j)^2 \left( \frac{5}{2}(\mathcal{Y}_R^j)^2 - \frac{2}{3} \Tr(\mathcal{Y}_R^2) \right) \overline{S^j_{\overline{\psi}\psi_R}}
	\, .
\end{split}\end{equation}
This counterterm action generates a BRST breaking,
\begin{equation}
\label{eq:BRSTtransfoSsct2}
\begin{split}
	\Delta_\text{sct}^2 =
	s_d{S_\text{sct}^2} =\;&
	\frac{-\hbar^2 e^4}{256 \pi^4} \frac{\Tr(\mathcal{Y}_R^4)}{6}
	\left( \frac{1}{\epsilon^2} - \frac{17}{12\epsilon} \right) \int \dInt[d]{x} (\overline{\partial}_\mu c) (\widehat{\partial}^2\bar{A}^\mu)
	\\
	& - \frac{\hbar^2 e^5}{256 \pi^4} \frac{1}{3\epsilon} \sum_{j}(\mathcal{Y}_R^j)^3\left( \frac{5}{2} (\mathcal{Y}_R^j)^2 - \frac{2}{3} \Tr(\mathcal{Y}_R^2) \right)
	\int \dInt[d]{x} c \, \overline{\partial}_\mu (\overline{\psi} \bar{\gamma}^\mu \Proj{R} \psi)
	\, ,
\end{split}
\end{equation}
where we have used the BRST invariance of $\overline{S_{AA}}$ and $S_{AA}$.
Again we note that, in contrast to the one-loop case, this BRST breaking
contains a non-evanescent contribution (the second term).
\section{BRST Symmetry Breaking and its Restoration at Two-Loop; Evaluation of the Two-Loop Finite Counterterm Action $S_\text{fct}^{(2)}$}
\label{subsect:2LoopBRSTRestore}

\newcommand{\GammaSubIDontLikeIt}{\Gamma_\text{sub}} 
\newcommand{\GammaSubDRegIDontLikeIt}{\Gamma_\text{sub}} 

This section is devoted to restoring BRST symmetry at the two-loop (or $\hbar^2$) order. We again follow the general procedure outlined
in \cref{sec:multiloopformulae} and proceed along the same
lines as at the one-loop level in \cref{sect:BRSTrestoration}.

At the two-loop level the structure of renormalization can be written
as
\begin{subequations}
\begin{align}
 \Gamma_\text{DReg}^{(2)} &= \Gamma^{(2)} + S_\text{sct}^2 + S_\text{fct}^2 \, ,
\label{eq:GammaDR2_2}
\\
	\Delta_\text{ct}^2 &= \mathcal{S}_d (S_0+S_\text{ct})^2 \, . 
\end{align}
\end{subequations}
The conditions (\ref{Ssctdefinition}, \ref{eq:SdGamExp4d}) on the counterterms specialize to
\begin{subequations}
\label{eq:deltadef}
\begin{align}
 S^2_\text{sct} +\Gamma^2_\text{div} &= 0 \, ,
\label{eq:S2sct_2L}
\\
\label{eq:CheckDeltasct2L} \big(\widehat{\Delta}\cdot\Gamma_\text{}^2 + \Delta_\text{ct}^1\cdot\Gamma_\text{}^{1}
+ \Delta_\text{ct}^2\big)_\text{div} &= 0 \, ,
\\
\text{LIM}_{d\to 4} \big(\widehat{\Delta}\cdot\Gamma_\text{}^2 + \Delta_\text{ct}^1\cdot\Gamma_\text{}^{1}
+ \Delta_\text{ct}^2\big)_\text{fin} &= 0 \, .
\end{align}
\end{subequations}
Like at the one-loop level, the second of these equations must hold
automatically and provides a consistency check; the third equation
determines the finite symmetry-restoring counterterms. We again
rewrite it as
\begin{align}
 N\big[\widehat{\Delta}\cdot\Gamma_\text{}^2 + \Delta_\text{ct}^1\cdot\Gamma_\text{}^{1}\big]
+ \Delta_\text{fct}^2 &= 0 \, ,
\label{eq:S2fct_2L}
\end{align}
which implicitly fixes the choice of finite, evanescent
counterterms. The meaning of this equation is as follows:  The breaking of the Slavnov-Taylor identity is given via the quantum action principle by Green functions with breaking insertions. The finite symmetry-restoring counterterms are defined such that they cancel the finite, purely 4-dimensional part of the breaking. As at the one-loop level, we have used that the BRST
variation of the singular counterterms $\Delta_\text{sct}^2$ contains
no terms of order $\epsilon^0$ and we could drop the index $\text{`DReg'}$.

In the following we first describe the required Feynman diagrammatic
computation, then we carry out the check corresponding to
\cref{eq:CheckDeltasct2L} and finally we determine the finite, symmetry-restoring
counterterms.

\subsection{Computation of the 2-Loop Breaking of BRST Symmetry}

Here we provide details on the computation of the
Feynman diagrams describing the two-loop symmetry breakings. As described in
\cref{sec:multiloopformulae} the quantum action principle implies that they
are given by diagrams with insertions of the symmetry breaking of the
tree-level and counterterm action. At the two-loop level
\cref{eq:CheckDeltasct2L,eq:S2fct_2L} show that the
divergent and finite parts of the following kinds of diagrams are
required. $\widehat{\Delta}\cdot\Gamma^2$ contains genuine two-loop
diagrams with one insertion of the tree-level breaking
$\widehat{\Delta}$, and it contains one-loop
diagrams with one insertion of a one-loop counterterm and one
insertion of $\widehat{\Delta}$. The object $\Delta_\text{ct}^1\cdot\Gamma^1$
consists of one-loop diagrams with tree-level vertices and with one
insertion of the breaking of the one-loop counterterm action
$\Delta_\text{ct}^1$. Note that in our case for the U(1) model we have $(\Delta^1_\text{sct}\cdot\Gamma^1)^2=0$, implying $(\Delta^1_\text{ct}\cdot\Gamma^1)^2=(\Delta^1_\text{fct}\cdot\Gamma^1)^2$, due to the fact that there are no ghost loop corrections.

\begin{figure}[t]
	\centering
	\begin{tabular}{*{3}{>{\centering\arraybackslash}m{0.3\textwidth}}}
		\includegraphics[scale=0.6]{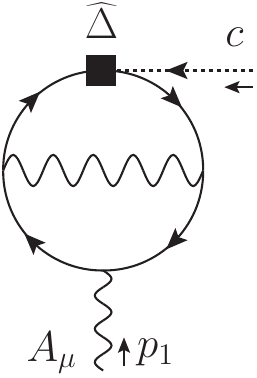}
		&
		\includegraphics[scale=0.6]{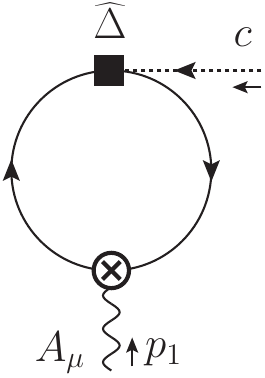}
		&
		\includegraphics[scale=0.6]{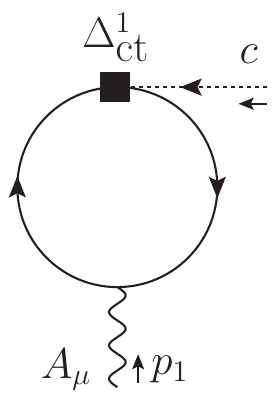}
	\end{tabular}
	\begin{tabular}{*{3}{>{\centering\arraybackslash}m{0.3\textwidth}}}
		\includegraphics[scale=0.6]{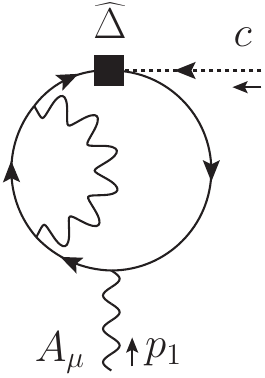} \newline
		+ loop on the other fermion propagator.
		&
		\includegraphics[scale=0.6]{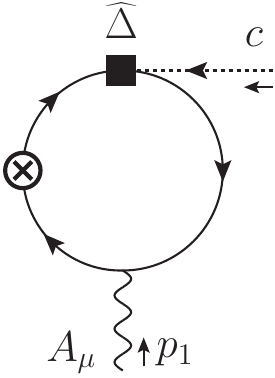} \newline
		+ fermion counterterm on the other fermion propagator.
		&
		\includegraphics[scale=0.6]{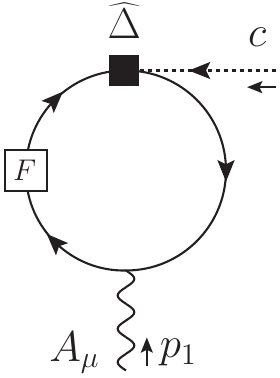} \newline
		+ fermion finite counterterm on the other fermion propagator.
	\end{tabular}
\caption{\label{fig:2LBRSTDGcA} 
  List of Feynman diagrams for the ghost--photon breaking contribution given in \cref{2LBRSTDGcA}.
The diagrams in the first column are genuine two-loop diagrams with one insertion of the tree-level breaking $\widehat{\Delta}$. The diagrams in the second column are one-loop diagrams with one insertion of a one-loop singular counterterm, denoted as a circled cross. The third column contains a one-loop diagram with an insertion of a one-loop symmetry-restoring counterterm, denoted by a boxed $F$, and a one-loop diagram with an insertion of the one-loop breaking $\Delta_\text{ct}^{1}$. }\end{figure}

\begin{figure}[t]
	\centering
	\begin{tabular}{*{3}{>{\centering\arraybackslash}m{0.3\textwidth}}}
		\includegraphics[scale=0.6]{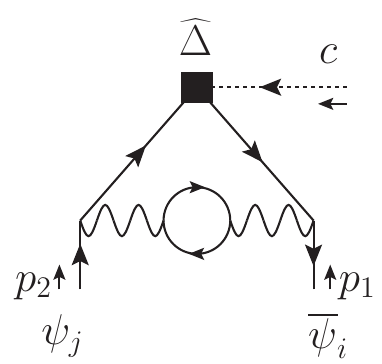}
		&
		\includegraphics[scale=0.6]{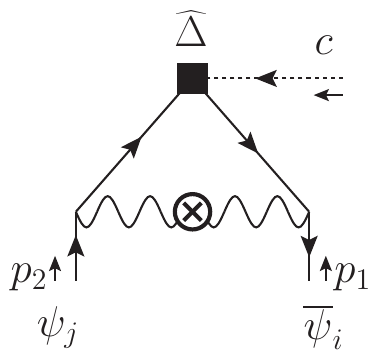}
		&
		\includegraphics[scale=0.6]{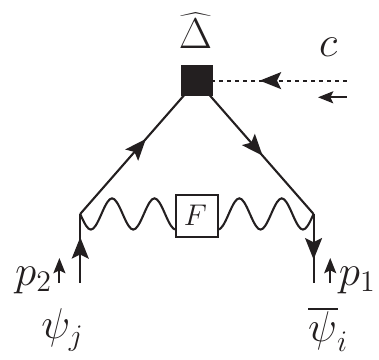}
	\end{tabular}
	\begin{tabular}{*{3}{>{\centering\arraybackslash}m{0.3\textwidth}}}
		\includegraphics[scale=0.6]{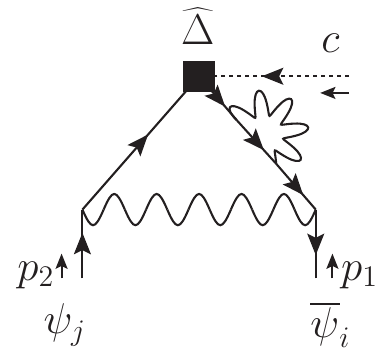} \newline
		+ loop on the other fermion propagator.
		&
		\includegraphics[scale=0.6]{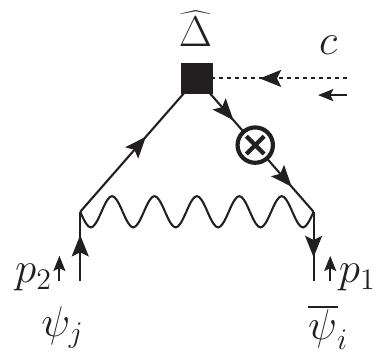} \newline
		+ fermion counterterm on the other fermion propagator.
		&
		\includegraphics[scale=0.6]{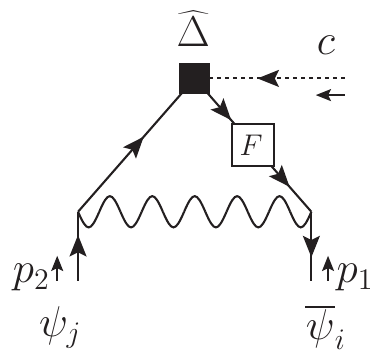} \newline
		+ fermion finite counterterm on the other fermion propagator.
	\end{tabular}
	\begin{tabular}{*{2}{>{\centering\arraybackslash}m{0.25\textwidth}}}
		\includegraphics[scale=0.6]{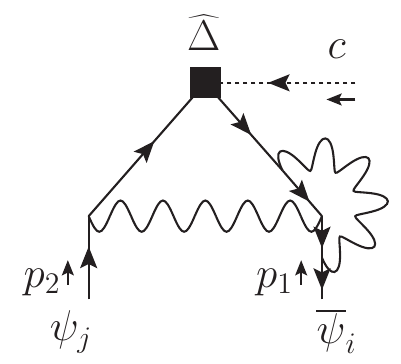} \newline
		+ loop on the other vertex.
		&
		\includegraphics[scale=0.6]{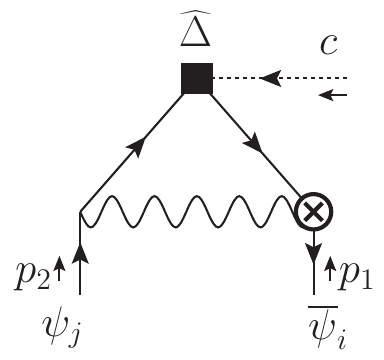} \newline
		+ fermion counterterm on the other vertex.
	\end{tabular}
	\begin{tabular}{*{3}{>{\centering\arraybackslash}m{0.3\textwidth}}}
		\includegraphics[scale=0.6]{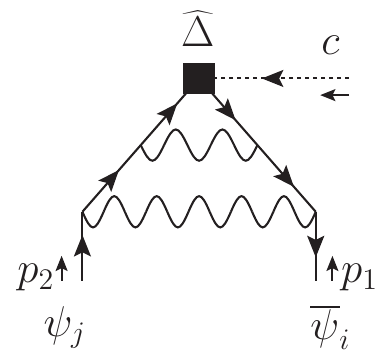}
		&
		\includegraphics[scale=0.6]{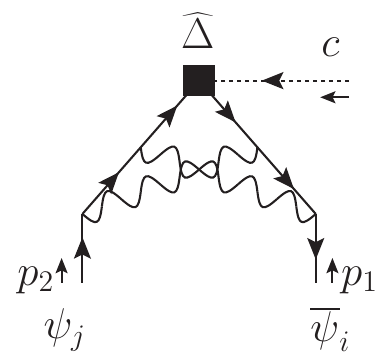}
		&
		\includegraphics[scale=0.6]{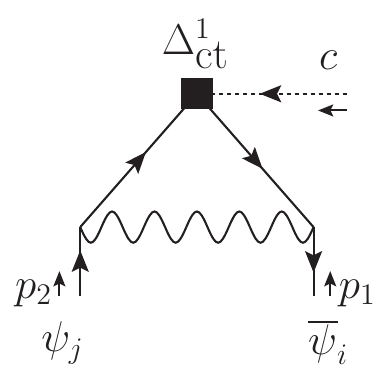}
	\end{tabular}

	\caption{\label{fig:2LBRSTDGpsibpsic}List of Feynman diagrams
          for the Ghost--fermion--fermion breaking contribution. The
          symbols are as in \cref{fig:2LBRSTDGcA} and the
          results are given in \cref{2LBRSTDGpsibpsic}.}
\end{figure}

\begin{figure}[t]
	\centering
	\begin{tabular}{*{3}{>{\centering\arraybackslash}m{0.3\textwidth}}}
		\includegraphics[scale=0.6]{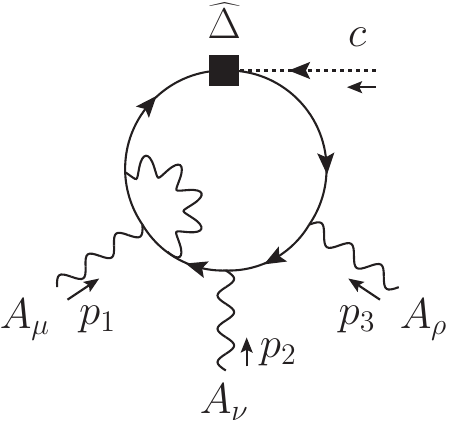} \newline
		+ loop on the other vertices.
		&
		\includegraphics[scale=0.6]{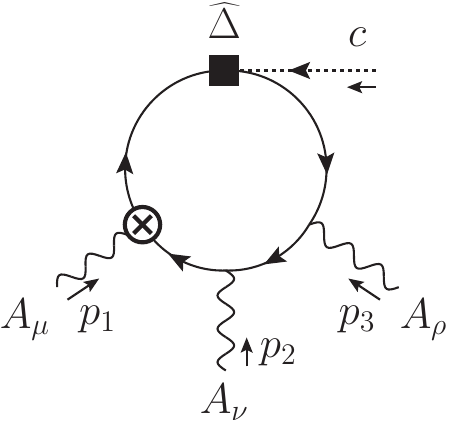} \newline
		+ fermion counterterm on the other vertices.
		&
		\raisebox{+33pt}{\includegraphics[scale=0.6]{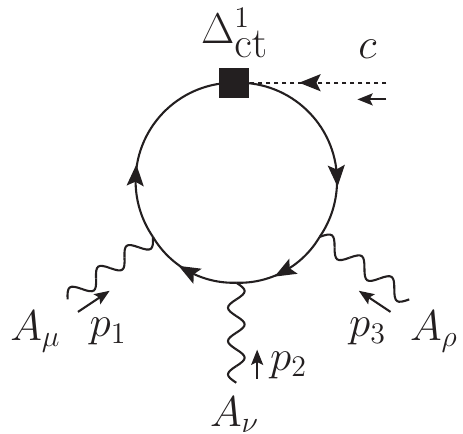}}
	\end{tabular}
	\begin{tabular}{*{2}{>{\centering\arraybackslash}m{0.3\textwidth}}}
		\includegraphics[scale=0.6]{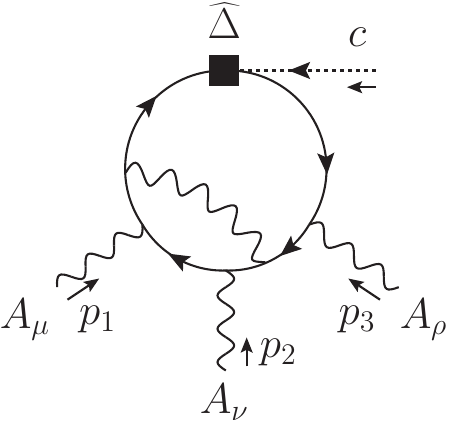} \newline
		+ mirrored (loop around $A_\nu$ and $A_\rho$ photons).
		&
		\raisebox{+30pt}{\includegraphics[scale=0.6]{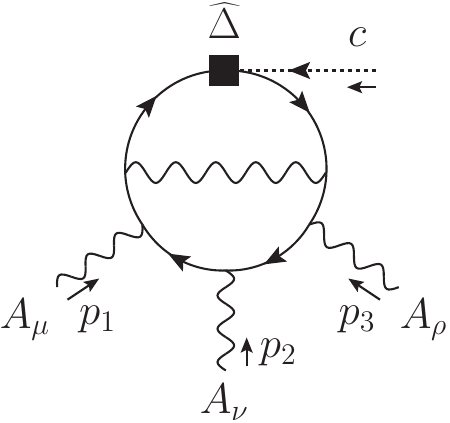}}
	\end{tabular}
	\begin{tabular}{*{3}{>{\centering\arraybackslash}m{0.3\textwidth}}}
		\raisebox{+15pt}{\includegraphics[scale=0.6]{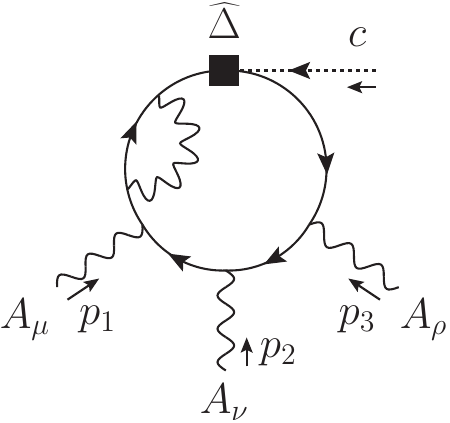}} \newline
		+ loop on the other fermion propagators.
		&
		\includegraphics[scale=0.6]{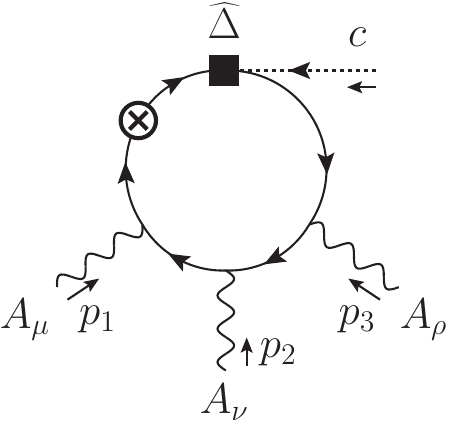} \newline
		+ fermion counterterm on the other fermion propagators.
		&
		\includegraphics[scale=0.6]{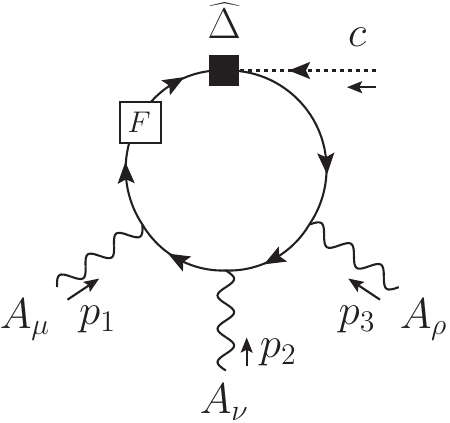} \newline
		+ fermion finite counterterm on the other fermion propagators.
	\end{tabular}

	\caption{\label{fig:2LBRSDGTcAAA}List of Feynman diagrams
          for the Ghost--three gauge bosons breaking contribution (additional diagrams corresponding to
          $\{ (p_1,\mu) \, , \, (p_2,\nu) \, , \, (p_3,\rho) \}$
          permutations are not shown).  The symbols are as in
          \cref{fig:2LBRSTDGcA} and the results are given in
          \cref{2LBRSDGTcAAA}.}
\end{figure}

Like at the one-loop level, the only relevant results are the ones
which are either divergent or finite but not evanescent. Since the
breaking insertions $\widehat{\Delta}$
are themselves evanescent, such results
can only arise from power-counting divergent Feynman diagrams.
For this reason only a finite number of Feynman diagrams with a specific
set of external fields need to be computed. The relevant diagrams with
non-vanishing contributions are
shown in \cref{fig:2LBRSTDGcA,fig:2LBRSTDGpsibpsic,fig:2LBRSDGTcAAA},
and their results are described in the following.


The ghost-gauge boson contribution from the diagrams with external fields
$cA$ shown in \cref{fig:2LBRSTDGcA} is
\begin{equation}
\label{2LBRSTDGcA}
	\imath \left( [\widehat{\Delta} + \Delta_\text{ct}^{1}] \cdot \widetilde{\Gamma} \right)^{2}_{A_\mu c} =
	\frac{1}{256 \pi^4} \frac{e^4 \Tr(\mathcal{Y}_R^4)}{6}
	\left[
	\left( \frac{1}{\epsilon^2} - \frac{17}{12\epsilon} \right) \hat{p}_1^2 \overline{p}_1^\mu
	- \frac{11}{4} \overline{p}_1^2 \overline{p}_1^\mu + \mathcal{O}(\hat{.})
	\right]
	\, .
\end{equation}
The result contains $1/\epsilon^2$ poles and $1/\epsilon$ poles with
local, evanescent coefficients and finite, non-evanescent
terms. Finite but evanescent terms are not relevant for
the present context and are suppressed here and in the following.

The ghost-fermion-fermion contribution from the diagrams with external fields
$c\overline{\psi}\psi$ shown in \cref{fig:2LBRSTDGpsibpsic} is
\begin{equation}
\label{2LBRSTDGpsibpsic}
\begin{split}
	\imath \left( [\widehat{\Delta} + \Delta_\text{ct}^{1}] \cdot \widetilde{\Gamma} \right)^{2}_{\psi \overline{\psi} c}
	=\;&
	\frac{1}{256 \pi^4} \frac{e^5 (\mathcal{Y}_R^j)^3}{3} (\bar{\slashed{p}}_1+\bar{\slashed{p}}_2) \Proj{R} \times\\
	& \left[
	\frac{1}{\epsilon}
	\left( \frac{5}{2} (\mathcal{Y}_R^j)^2 - \frac{2}{3} \Tr(\mathcal{Y}_R^2) \right)
	+ \frac{127}{12} (\mathcal{Y}_R^j)^2 - \frac{1}{9} \Tr(\mathcal{Y}_R^2)
	\right]
	\, .
\end{split}
\end{equation}
The result contains no $1/\epsilon^2$ poles but only $1/\epsilon$ poles with
local, evanescent coefficients and finite, non-evanescent
terms.

The ghost-two gauge bosons contribution from diagrams with external fields
$cAA$ turns out to vanish. Hence
\begin{equation}
\label{2LBRSTDGcAA}
	\imath \left( [\widehat{\Delta} + \Delta_\text{ct}^{1}] \cdot \widetilde{\Gamma} \right)^{2}_{AAc}
	= 0
	\, .
\end{equation}

The ghost-three gauge bosons contribution from the diagrams with external
fields $cAAA$ shown in \cref{fig:2LBRSDGTcAAA} is
\begin{equation}
\label{2LBRSDGTcAAA}
\begin{split}
	\imath \left( [\widehat{\Delta} + \Delta_\text{ct}^{1}] \cdot \widetilde{\Gamma} \right)^{2}_{A_\rho A_\nu A_\mu c}
	=
	\frac{1}{256 \pi^4} 3 e^6 \Tr(\mathcal{Y}_R^6)
	(\overline{p}_1 + \overline{p}_2 + \overline{p}_3)_\sigma
	\\
	( \bar{g}^{\mu\nu} \bar{g}^{\rho\sigma} + \bar{g}^{\mu\rho} \bar{g}^{\nu\sigma} + \bar{g}^{\mu\sigma} \bar{g}^{\nu\rho} )
	\, .
\end{split}
\end{equation}
Notice that this result contains no UV divergence but only finite terms.

Collecting the results of
\cref{2LBRSTDGcA,2LBRSTDGpsibpsic,2LBRSTDGcAA,2LBRSDGTcAAA}, one
obtains the following result for the two-loop breaking of the
Slavnov-Taylor identity of two-loop subrenormalized Green functions:
\begin{equation}
\begin{split}
 \left( [\widehat{\Delta} + \Delta_\text{ct}^{1}] \cdot \Gamma \right)^{2} &=
 \frac{\hbar^2 e^4}{256 \pi^4}
 \int \dInt[d]{x} \\
	& \, \bigg\{
	 -\frac{\Tr(\mathcal{Y}_R^4)}{6}
	\left[
	\left( \frac{1}{\epsilon^2} -\frac{17}{12\epsilon}\right)
	c\, \overline{\partial}_\mu \widehat{\partial}^2 \bar{A}^\mu
	- \frac{11}{4} c\, \overline{\partial}_\mu \overline{\partial}^2 \bar{A}^\mu
	\right]
	\\
	&+ e \sum_{j}\frac{(\mathcal{Y}_R^j)^3}{3}
	\bigg[
	\frac{1}{\epsilon}
	\left( \frac{5}{2} (\mathcal{Y}_R^j)^2 - \frac{2}{3} \Tr(\mathcal{Y}_R^2) \right)
	\\
	&
	+ \frac{127}{12} (\mathcal{Y}_R^j)^2 - \frac{1}{9} \Tr(\mathcal{Y}_R^2) \bigg]
	c\, \overline{\partial}_\mu (\overline{\psi}_j \bar{\gamma}^\mu \,\Proj{R}\, \psi_j)\,
	\\
	&+\frac{3\,e^2\Tr(\mathcal{Y}_R^6)}{2}
	c\, \overline{\partial}_\mu ( \bar{A}^\mu \bar{A}_\nu \bar{A}^\nu )\bigg\}+\mathcal{O}(\hat{.})
	\, .
\end{split}
\end{equation}
It is particularly noteworthy that, despite significantly more
complicated computations, the structure of the terms is the same as at
the one-loop level.

\subsection{Two-loop Finite Symmetry-Restoring Counterterms}

Like at the one-loop level we can first use the result to check the
cancellation of the UV divergences as prescribed by
\cref{eq:CheckDeltasct2L}. Indeed, this cancellation  with
$s_d{S^{(2)}_\text{sct}}$, \cref{eq:BRSTtransfoSsct2} takes place as
it should, in the explicit form
\begin{equation}
\label{eq:sct2cancellation}
\Delta_\text{sct}^2 = s_d{S^{(2)}_\text{sct}}
= - \left( [\widehat{\Delta} + \Delta_\text{ct}^{1}] \cdot \widetilde{\Gamma} \right)^{2}_\text{div} \, ,
\end{equation}
providing a confirmation of the computation.
Next we can turn to the determination of the two-loop
symmetry-restoring counterterms, using \cref{eq:S2fct_2L}.
Given the results of the previous subsection and a simple calculation we obtain
%
\begin{equation}\label{eq:BRSTtransfoSfct2}\begin{split}
\Delta_\text{fct}^2
=\;&
-
N\big[\widehat{\Delta}\cdot\Gamma_\text{DReg}^2 +
\Delta_\text{ct}^1\cdot\Gamma_\text{DReg}^{1}\big]
\\
=\;&-\mathop{\text{LIM}}_{d \to 4} \left\{ \left( [\widehat{\Delta} + \Delta_\text{ct}^{(1)}] \cdot {\Gamma} \right)^{(2)} + s_d{S^{(2)}_\text{sct}} \right\}
\\
=\;&
+ \frac{\hbar^2 e^4}{256 \pi^4} \Tr(\mathcal{Y}_R^4) \; s\left( \frac{11}{48} \int \dInt[4]{x} \bar{A}_\mu \overline{\partial}^2 \bar{A}^\mu \right) \\
&- \frac{\hbar^2 e^4}{256 \pi^4} \sum_{j}(\mathcal{Y}_R^j)^2\,
\left( \frac{127}{36} (\mathcal{Y}_R^j)^2 - \frac{1}{27} \Tr(\mathcal{Y}_R^2) \right)
s \int \dInt[4]{x} \overline{\psi}_j \imath \bar{\slashed{\partial}} \,\Proj{R}\, \psi_j \\
&+ \frac{\hbar^2 e^6}{256 \pi^4} \frac{3\,\Tr(\mathcal{Y}_R^6)}{8} \;
s \int \dInt[4]{x} \bar{A}_\mu \bar{A}^\mu \bar{A}_\nu \bar{A}^\nu
\, .
\end{split}\end{equation}
Here the right-most equation has been written as an explicit
4-dimensional BRST transformation of a local action.

This implies that the following choice of finite counterterms restores
the Slavnov-Taylor identity at the two-loop level,
\begin{equation}
\label{eq:Sfct2L}
\begin{split}
S^{2}_\text{fct} =\;&
\left(\frac{\hbar}{16\pi^2}\right)^2 \int \dInt[4]{x}
e^4 \left\{
 \Tr(\mathcal{Y}_R^4) \frac{11}{48} \bar{A}_\mu \overline{\partial}^2 \bar{A}^\mu
+ 3 \,e^2 \frac{\Tr(\mathcal{Y}_R^6)}{8}\, \bar{A}_\mu \bar{A}^\mu \bar{A}_\nu \bar{A}^\nu
\right.
\\
&\left.
- \sum_{j}(\mathcal{Y}_R^j)^2 \,
\left( \frac{127}{36} (\mathcal{Y}_R^j)^2 - \frac{1}{27} \Tr(\mathcal{Y}_R^2) \right)
\Big( \overline{\psi}_j \imath \bar{\slashed{\partial}} \,\Proj{R}\, \psi_j \Big)
\right\}
\, .
\end{split}\end{equation}
Like at the one-loop level, three kinds of terms exist. In an obvious
way they correspond to the restoration of the Ward identity relations
for the photon self energy, the photon 4-point function and the
fermion self energy/photon interaction.

\subsection{Tests of Ward Identities}

In order to check the consistency of the previously calculated
divergent and finite counterterms we may make use of Ward identities
which express relations of Green's functions and their properties due
to gauge invariance of the theory. In \cref{sect:definingsymmetries}
we have seen that in our $U(1)$ model the Slavnov-Taylor identity
straightforwardly leads to Ward identities since certain functional
relations trivially survive renormalization. Once the Slavnov-Taylor
identity is satisfied, the Ward identities will likewise be valid, but
they provide a check that is independent of breaking
diagrams. \cref{eq:GeneralWI} supplies us with three well-known QED
Ward identities for renormalized Green functions to check our
counterterm results:
\begin{enumerate}
	\item The transversality of the photon self energy,
	\begin{equation}
	\label{eq:Ward1}
	\imath p_\nu \frac{\delta^2{\widetilde{\Gamma}_\text{ren}}}{\delta{A_\mu(p)} \delta{A_\nu(-p)}} = 0 \, ;
	\end{equation}

	\item The transversality of multi-photon vertices, and in particular the
	photon 4-point amplitude,
	\begin{equation}
	\label{eq:Ward2}
	\imath (p_{1+2+3})_\sigma \frac{\delta^4{\widetilde{\Gamma}_\text{ren}}}{\delta{A_\rho(p_3)} \delta{A_\nu(p_2)} \delta{A_\mu(p_1)} \delta{A_\sigma(-p_{1+2+3})}} = 0
	\end{equation}
	(denoting $p_{1+2+3} \equiv p_1+p_2+p_3$);

	\item The relation between fermion self energy and fermion-photon interaction for vanishing photon momentum $q=0$,
	\begin{equation}
	\label{eq:Ward3}
	-ie\mathcal{Y}_R\frac{\partial}{\partial p_{\mu}}\frac{\delta^2\widetilde{\Gamma}_\text{ren}}{\delta{\overline{\psi}(-p)} \delta{\psi(p)}}+\imath\,\frac{\delta^3\widetilde{\Gamma}_\text{ren}}{\delta{A_{\mu}(0)} \delta{\overline{\psi}(-p)} \delta{\psi(p)}} = 0 \, .
	\end{equation}
\end{enumerate}
With these equations, we can demonstrate the consistency and correctness of our calculations by evaluating usual loop diagrams and compare them with the results for breaking insertions.

We begin with the example of the two-loop divergent part of the photon self energy. If we contract it with one momentum, what we obtain is
\begin{equation}
    \left. \imath \, p_{\nu} \, \widetilde{\Gamma}_{A(-p)A(p)}^{\mu\nu} \right|^{2}_\text{div} =
    \frac{\imath e^4}{256 \pi^4} \frac{\Tr(\mathcal{Y}_R^4)}{6}  \left( \frac{17}{12 \epsilon} - \frac{1}{\epsilon^2} \right) \widehat{p}^2 \overline{p}^{\mu} = -\left( [\widehat{\Delta} + \Delta_\text{ct}^{1}] \cdot \widetilde{\Gamma} \right)^{2}_{\text{div},\, A_{\mu}(-p)c(p)}
    \, .
\end{equation}
The first of these equations is obtained by direct computation of the
appropriate two-loop diagrams. The second equation is then an
observation using \cref{eq:BRSTtransfoSsct2} and \cref{eq:sct2cancellation}.
These equations mean that the part of the divergent photon self energy
that would violate transversality is cancelled by the divergent
counterterm calculated from the breaking insertion.

The finite part of photon self energy at the two loop level is given by
\begin{equation}
    \left. \imath \widetilde{\Gamma}_{AA}^{\mu\nu}(p) \right|^{2}_\text{fin} =
    \frac{\imath e^4}{256 \pi^4} \frac{\Tr(\mathcal{Y}_R^4)}{3} \left[ \left( \frac{673}{23} -6 \log(-\overline{p}^2)-24\zeta(3) \right) (\overline{p}^\mu \overline{p}^\nu - \overline{p}^2 \overline{g}^{\mu\nu}) + \frac{11}{8} \overline{p}^\mu \overline{p}^\nu \right]
    \, ,
\end{equation}
and after the momentum contraction we obtain
\begin{equation}
    \left. \imath \, p_{\nu} \, \widetilde{\Gamma}_{A(-p)A(p)}^{\mu\nu} \right|^{2}_\text{fin} =
    \frac{\imath e^4}{256 \pi^4} \frac{\Tr(\mathcal{Y}_R^4)}{6}  \frac{11}{4}  \overline{p}^2 \overline{p}^{\mu} = -\left( [\widehat{\Delta} + \Delta_\text{ct}^{1}] \cdot \widetilde{\Gamma} \right)^{2}_{\text{fin},\, A_{\mu}(-p)c(p)}
    \, .
\end{equation}
The first of these equations is again obtained by direct computation
of the diagrams. It illustrates that the non-local $\log(-\overline{p}^2)$
and transcendental $\zeta(3)$ parts are by
themselves transversal. The second equation is then observed by
comparison with \cref{eq:BRSTtransfoSfct2}. Hence we confirm
that the violation of the symmetry is  restored by our finite
counterterm evaluated from breaking diagrams.

The 4-photon amplitude is finite.
A direct, explicit  manipulation of the corresponding Feynman diagrams
shows that we can relate the breaking of the Ward identity to the
breaking of the Slavnov-Taylor identity as
\begin{equation}
\label{eq:FiniteWard4A}
\begin{split}
    &- \left. \imath \, p_{\nu} \, \widetilde{\Gamma}_{A(-p_1) A(-p_2) A(-p_3) A(p)}^{\mu_1 \mu_2 \mu_3 \nu} \right|^{2}_\text{fin} =
    \left( [\widehat{\Delta} + \Delta_\text{ct}^{1}] \cdot \widetilde{\Gamma} \right)_{\text{fin},\, A_{\mu_1}(-p_1) A_{\mu_2}(-p_2) A_{\mu_3}(-p_3) c(p)} \\
    &= \frac{\imath e^6}{256 \pi^4}\, 3 \Tr(\mathcal{Y}_R^6)\, (\overline{p}_1 + \overline{p}_2 + \overline{p}_3)_\nu (\overline{g}^{\nu\mu_1} \overline{g}^{\mu_2\mu_3} + \overline{g}^{\nu\mu_2} \overline{g}^{\mu_1\mu_3} + \overline{g}^{\nu\mu_3} \overline{g}^{\mu_1\mu_2}) \, .
\end{split}
\end{equation}
Via \cref{eq:BRSTtransfoSfct2} this shows again that the counterterms
of \cref{eq:Sfct2L} appropriately restore this Ward identity.

We can investigate the Ward identity between the fermion self energy
and fermion-photon interaction \cref{eq:Ward3} in a similar way.
The divergent two-loop violation is given by
\begin{equation}
\label{eq:DivWard3}
\begin{split}
    &- \left. \imath\, e \mathcal{Y}_R \frac{\partial}{\partial p_\mu}\widetilde{\Gamma}_{\psi(p) \overline{\psi}(-p)} \right|^{2}_\text{div} + \left. \imath\,\widetilde{\Gamma}_{\psi(p) \overline{\psi}(-p) A(0)}^\mu \right|^{2}_\text{div} \\
    &= \imath\mathcal{Y}_R \frac{ e^5}{256 \pi^4 \epsilon}  \overline{\gamma}^\mu\, \Proj{R} \bigg(\frac{2\mathcal{Y}_R^2\,\Tr(\mathcal{Y}_R^2)}{9}-\frac{5\mathcal{Y}_R^4}{6} \,
    \bigg)\\
    &=-\frac{\partial}{\partial q_{\mu}}\left( [\widehat{\Delta} + \Delta_\text{ct}^{1}] \cdot \widetilde{\Gamma} \right)^{2}_{\text{div},\, \psi(p-q)\overline{\psi}(-p)c(q)}(q=0) \, ,
\end{split}
\end{equation}
and the finite two-loop violation by
\begin{equation}
\label{eq:FiniteWard3}
\begin{split}
    &- \left. \imath\, e \mathcal{Y}_R \frac{\partial}{\partial p_\mu}\widetilde{\Gamma}_{\psi(p) \overline{\psi}(-p)} \right|^{2}_\text{fin} + \left. \imath\,\widetilde{\Gamma}_{\psi(p) \overline{\psi}(-p) A(0)}^\mu \right|^{2}_\text{fin} \\
    &=\imath\mathcal{Y}_R \frac{ e^5}{256 \pi^4 }  \overline{\gamma}^\mu\, \Proj{R} \bigg(\log(-\overline{p}^2)\bigg(\frac32\mathcal{Y}_R^4-\mathcal{Y}_R^2\Tr(\mathcal{Y}_R^2) \bigg)+\frac{62}{27}\mathcal{Y}_R^2\Tr(\mathcal{Y}_R^2)-\frac{109}{72}\mathcal{Y}_R^4\\
    &\qquad \qquad\qquad \quad \quad - \log(-\overline{p}^2)\bigg(\frac32\mathcal{Y}_R^4-\mathcal{Y}_R^2\Tr(\mathcal{Y}_R^2) \bigg)
    -\frac{61}{27}\mathcal{Y}_R^2\Tr(\mathcal{Y}_R^2)-\frac{145}{72}\mathcal{Y}_R^4\bigg)\\
    &= \imath\mathcal{Y}_R \frac{ e^5}{256 \pi^4 }  \overline{\gamma}^\mu\, \Proj{R} \bigg(\frac{\mathcal{Y}_R^2\,\Tr(\mathcal{Y}_R^2)}{27}-\frac{127\mathcal{Y}_R^4}{36} \,
    \bigg)\\
    &=-\frac{\partial}{\partial q_{\mu}}\left( [\widehat{\Delta} + \Delta_\text{ct}^{1}] \cdot \widetilde{\Gamma} \right)^{2}_{\text{fin},\, \psi(p-q)\overline{\psi}(-p)c(q)}(q=0) \, .
\end{split}
\end{equation}
In each case again the first equations are obtained from explicit
computation of the Feynman diagrams, and the last equations are
obtained by comparing with \cref{eq:BRSTtransfoSsct2}, \cref{eq:sct2cancellation} and
\cref{eq:BRSTtransfoSfct2}.\footnote{%
	The divergent $1/\epsilon^2$ poles in \eqref{eq:DivWard3} are
	omitted since they cancel completely. The second and third rows in
	\eqref{eq:FiniteWard3} represent the full results for finite
	(momentum-differentiated) photon self energy and vertex interaction,
	respectively.}
As a result, it is established that the
counterterms in  \cref{eq:Sfct2L}  restore all Ward identities.

\section{Conclusions}
\label{sect:Concl}

In this work we applied the BMHV scheme for non-anticommuting $\gamma_5$ in dimensional
regularization to a chiral gauge theory at the two-loop level, and we
studied the BMHV-specific aspects of renormalization. Most importantly
we determined the full structure of two-loop symmetry-restoring
counterterms. The present work is restricted to an abelian gauge
theory with right-handed fermions and establishes the methodology. The
same method will be applicable to general non-abelian gauge theories
with scalar and fermionic matter.

In general, the application of the BMHV scheme leads to several
specific kinds of counterterms: the
ultraviolet (UV) divergences cannot be cancelled by counterterms
generated by field and parameter renormalization; additional,
UV divergent evanescent counterterms (corresponding to operators which
vanish in strictly 4 dimensions) are needed; and the breaking of BRST
symmetry needs to be repaired by adding finite, symmetry-restoring
counterterms. We have evaluated all these counterterms explicitly at the one-loop
and two-loop level.
An important aspect of our results is that the structure at the
one-loop and two-loop level is essentially the same.
As expected, the UV divergences arise in the fermion and the photon self energy
and in the fermion--photon interaction. The triple and quartic photon self
interactions are UV finite. However, there are purely evanescent divergences in
the photon self energy, and at the two-loop level there is a
non-evanescent divergence in the fermion self energy, both of which
require an extra counterterm which cannot be obtained from field or
parameter renormalization.

The required symmetry-restoring counterterms turn out to take a rather
simple structure with a straightforward interpretation. Both at the
one-loop and the two-loop level there are three kinds of such
counterterms. A counterterm to the photon self energy
restores transversality of the renormalized photon self energy.
Similarly, a counterterm to the photon 4-point function
restores the Ward identity for this Green function. Finally, a
counterterm to the fermion self energy restores its Ward identity-like
relation to the fermion--photon interaction. An important outcome is
that the precise form of these counterterms is now known, and it is
established that this is the complete set of symmetry-restoring
counterterms for arbitrary two-loop calculations in the model.

We applied a method which was previously applied at the one-loop level
in Refs.\ \cite{Martin:1999cc,Belusca-Maito:2020ala};
Refs.\ \cite{Stockinger:2005gx,Hollik:2005nn,Stockinger:2018oxe}
applied similar techniques at the
multiloop level, however in cases where the symmetry is actually
unbroken by the regularization. The core of the method is the
evaluation of the breaking of the Slavnov-Taylor identity by employing
the regularized quantum action principle
\cite{Breitenlohner:1975qe}. Here we presented the first such
computation at the two-loop level. It involves Feynman diagrams of
four different kinds: genuine two-loop diagrams with an insertion of
the tree-level breaking $\widehat{\Delta}$ and one-loop diagrams with
insertions of the one-loop breaking $\Delta_\text{ct}^1$ or of the
one-loop divergent or finite counterterms.

Since the method is now established and not restricted to abelian
theories, it will be possible to apply it to general non-abelian
chiral gauge theories and to the Standard Model at the two-loop
level. In this way, two-loop Standard Model calculations will become
feasible in the BMHV scheme without worrying about symmetry
violations or scheme inconsistencies. As a further outlook, it will be
of interest to explore in detail the relationship between the modified
counterterm structure (with additional UV divergent and non-symmetric
finite terms) and the renormalization group, similar to the one-loop
discussion of Ref.\ \cite{Belusca-Maito:2020ala}.

\acknowledgments

The authors highly acknowledge the financial support from the Croatian Science Foundation (HRZZ) under the project ``PRECIOUS'' (``Precise Computations of Physical Observables in Supersymmetric Models'') number \verb|HRZZ-IP-2016-06-7460|. In addition A.I. acknowledges financial support of the previous Croatian Science Foundation project \\ \verb|HRZZ-IP-2013-11-8799|. P.K. and D.S. acknowledge financial support by the German Science
Foundation DFG, grant STO 876/8-1.
We also thank P.\ Marquard for discussions of 2-loop techniques and
Refs. \cite{Misiak:1994zw} and \cite{Chetyrkin:1997fm}.

\newpage

\bibliographystyle{JHEPmod}
\bibliography{ChiralQEDv2}

\providecommand{\href}[2]{#2}\begingroup\raggedright\begin{thebibliography}{10}

\bibitem{Cicuta:1972jf}
G.~Cicuta and E.~Montaldi, \emph{{Analytic renormalization via continuous space
  dimension}}, \href{https://doi.org/10.1007/BF02756527}{\emph{Lett. Nuovo
  Cim.} {\bfseries 4} (1972) 329}.

\bibitem{Bollini:1972ui}
C.~Bollini and J.~Giambiagi, \emph{{Dimensional Renormalization: The Number of
  Dimensions as a Regularizing Parameter}},
  \href{https://doi.org/10.1007/BF02895558}{\emph{Nuovo Cim. B} {\bfseries 12}
  (1972) 20}.

\bibitem{Ashmore:1972uj}
J.~Ashmore, \emph{{A Method of Gauge Invariant Regularization}},
  \href{https://doi.org/10.1007/BF02824407}{\emph{Lett. Nuovo Cim.} {\bfseries
  4} (1972) 289}.

\bibitem{tHooft:1972tcz}
G.~'t~Hooft and M.~J.~G. Veltman, \emph{{Regularization and Renormalization of
  Gauge Fields}},
  \href{https://doi.org/10.1016/0550-3213(72)90279-9}{\emph{Nucl. Phys.}
  {\bfseries B44} (1972) 189}.

\bibitem{Chanowitz:1979zu}
M.~S. Chanowitz, M.~Furman and I.~Hinchliffe, \emph{{The Axial Current in
  Dimensional Regularization}},
  \href{https://doi.org/10.1016/0550-3213(79)90333-X}{\emph{Nucl. Phys.}
  {\bfseries B159} (1979) 225}.

\bibitem{Kreimer:1989ke}
D.~Kreimer, \emph{{The $\gamma_5$ Problem and Anomalies: A Clifford Algebra
  Approach}}, \href{https://doi.org/10.1016/0370-2693(90)90461-E}{\emph{Phys.
  Lett.} {\bfseries B237} (1990) 59}.

\bibitem{Korner:1991sx}
J.~G. Korner, D.~Kreimer and K.~Schilcher, \emph{{A Practicable $\gamma_5$
  scheme in dimensional regularization}},
  \href{https://doi.org/10.1007/BF01559471}{\emph{Z. Phys.} {\bfseries C54}
  (1992) 503}.

\bibitem{Kreimer:1993bh}
D.~Kreimer, \emph{{The Role of $\gamma_5$ in dimensional regularization}},
  \href{https://arxiv.org/abs/hep-ph/9401354}{{\ttfamily hep-ph/9401354}}
  UTAS-PHYS-94-01, 1993,
  [\href{https://arxiv.org/abs/hep-ph/9401354}{{\ttfamily hep-ph/9401354}}].

\bibitem{Larin:1993tq}
S.~A. Larin, \emph{{The Renormalization of the axial anomaly in dimensional
  regularization}},
  \href{https://doi.org/10.1016/0370-2693(93)90053-K}{\emph{Phys. Lett.}
  {\bfseries B303} (1993) 113}.

\bibitem{Trueman:1995ca}
T.~L. Trueman, \emph{{Spurious anomalies in dimensional renormalization}},
  \href{https://doi.org/10.1007/BF02907437}{\emph{Z. Phys.} {\bfseries C69}
  (1996) 525}.

\bibitem{Chetyrkin:1997gb}
K.~G. Chetyrkin, M.~Misiak and M.~Munz, \emph{{$|\Delta F| = 1$ nonleptonic
  effective Hamiltonian in a simpler scheme}},
  \href{https://doi.org/10.1016/S0550-3213(98)00131-X}{\emph{Nucl. Phys.}
  {\bfseries B520} (1998) 279}.

\bibitem{Jegerlehner:2000dz}
F.~Jegerlehner, \emph{{Facts of life with $\gamma_5$}},
  \href{https://doi.org/10.1007/s100520100573}{\emph{Eur. Phys. J.} {\bfseries
  C18} (2001) 673}.

\bibitem{Bednyakov:2015ooa}
A.~V. Bednyakov and A.~F. Pikelner, \emph{{Four-loop strong coupling
  beta-function in the Standard Model}},
  \href{https://doi.org/10.1016/j.physletb.2016.09.007}{\emph{Phys. Lett.}
  {\bfseries B762} (2016) 151}
  [\href{https://arxiv.org/abs/1508.02680}{{\ttfamily 1508.02680}}].

\bibitem{Zoller:2015tha}
M.~F. Zoller, \emph{{Top-Yukawa effects on the $\beta$-function of the strong
  coupling in the SM at four-loop level}},
  \href{https://doi.org/10.1007/JHEP02(2016)095}{\emph{JHEP} {\bfseries 02}
  (2016) 095}.

\bibitem{Bruque:2018bmy}
A.~M. Bruque, A.~L. Cherchiglia and M.~Pérez-Victoria, \emph{{Dimensional
  regularization vs methods in fixed dimension with and without $\gamma_5$}},
  \href{https://doi.org/10.1007/JHEP08(2018)109}{\emph{JHEP} {\bfseries 08}
  (2018) 109}.

\bibitem{Gnendiger:2017rfh}
C.~Gnendiger and A.~Signer, \emph{{$\gamma_5$ in the four-dimensional helicity
  scheme}}, \href{https://doi.org/10.1103/PhysRevD.97.096006}{\emph{Phys. Rev.
  D} {\bfseries 97} (2018) 096006}
  [\href{https://arxiv.org/abs/1710.09231}{{\ttfamily 1710.09231}}].

\bibitem{Poole:2019kcm}
C.~Poole and A.~E. Thomsen, \emph{{Constraints on 3- and 4-loop
  $\beta$-functions in a general four-dimensional Quantum Field Theory}},
  \href{https://doi.org/10.1007/JHEP09(2019)055}{\emph{JHEP} {\bfseries 09}
  (2019) 055} [\href{https://arxiv.org/abs/1906.04625}{{\ttfamily
  1906.04625}}].

\bibitem{Poole:2019txl}
C.~Poole and A.~E. Thomsen, \emph{{Weyl Consistency Conditions and
  $\gamma_5$}},
  \href{https://doi.org/10.1103/PhysRevLett.123.041602}{\emph{Phys. Rev. Lett.}
  {\bfseries 123} (2019) 041602}
  [\href{https://arxiv.org/abs/1901.02749}{{\ttfamily 1901.02749}}].

\bibitem{Zerf:2019ynn}
N.~Zerf, \emph{{Fermion Traces Without Evanescence}},
  \href{https://doi.org/10.1103/PhysRevD.101.036002}{\emph{Phys. Rev.}
  {\bfseries D101} (2020) 036002}.

\bibitem{Ahmed:2020kme}
T.~Ahmed, W.~Bernreuther, L.~Chen and M.~Czakon, \emph{{Polarized $q \bar{q}
  \rightarrow Z +$Higgs amplitudes at two loops in QCD: the interplay between
  vector and axial vector form factors and a pitfall in applying a
  non-anticommuting $\gamma_5$}},
  \href{https://doi.org/10.1007/JHEP07(2020)159}{\emph{JHEP} {\bfseries 07}
  (2020) 159}.

\bibitem{Ahmed:2021spj}
T.~Ahmed, L.~Chen and M.~Czakon, \emph{{Renormalization of the flavor-singlet
  axial-vector current and its anomaly in dimensional regularization}},
  \href{https://doi.org/10.1007/JHEP05(2021)087}{\emph{JHEP} {\bfseries 05}
  (2021) 087}.

\bibitem{Cherchiglia:2020iug}
A.~Cherchiglia, D.~C. Arias-Perdomo, A.~R. Vieira, M.~Sampaio and B.~Hiller,
  \emph{{Two-loop renormalisation of gauge theories in $4D$ Implicit
  Regularisation: transition rules to dimensional methods}},
  \href{https://doi.org/10.1140/epjc/s10052-021-09259-6}{\emph{Eur. Phys. J. C}
  {\bfseries 81} (2021) 468}
  [\href{https://arxiv.org/abs/2006.10951}{{\ttfamily 2006.10951}}].

\bibitem{TorresBobadilla:2020ekr}
W.~J. Torres~Bobadilla et~al., \emph{{May the four be with you: Novel
  IR-subtraction methods to tackle NNLO calculations}},
  \href{https://doi.org/10.1140/epjc/s10052-021-08996-y}{\emph{Eur. Phys. J. C}
  {\bfseries 81} (2021) 250}.

\bibitem{Cherchiglia:2021uce}
A.~Cherchiglia, \emph{{A step towards a consistent treatment of chiral theories
  at higher loop order: the abelian case}},
  \href{https://arxiv.org/abs/2106.14039}{{\ttfamily 2106.14039}}.

\bibitem{Cherchiglia:2021yxz}
A.~Cherchiglia, \emph{{Two-loop gauge coupling $\beta$-function in a
  four-dimensional framework: the Standard Model case}},  in \emph{{15th
  International Symposium on Radiative Corrections: Applications of Quantum
  Field Theory to Phenomenology AND LoopFest XIX: Workshop on Radiative
  Corrections for the LHC and Future Colliders}}, 10, 2021,
  \href{https://arxiv.org/abs/2110.01739}{{\ttfamily 2110.01739}}.

\bibitem{Akyeampong:1973xi}
D.~Akyeampong and R.~Delbourgo, \emph{{Dimensional regularization, abnormal
  amplitudes and anomalies}},
  \href{https://doi.org/10.1007/BF02786835}{\emph{Nuovo Cim. A} {\bfseries 17}
  (1973) 578}.

\bibitem{Akyeampong:1973vk}
D.~Akyeampong and R.~Delbourgo, \emph{{Dimensional regularization and PCAC}},
  \href{https://doi.org/10.1007/BF02820839}{\emph{Nuovo Cim. A} {\bfseries 18}
  (1973) 94}.

\bibitem{Akyeampong:1973vj}
D.~Akyeampong and R.~Delbourgo, \emph{{Anomalies via dimensional
  regularization}}, \href{https://doi.org/10.1007/BF02801848}{\emph{Nuovo Cim.
  A} {\bfseries 19} (1974) 219}.

\bibitem{Breitenlohner:1975qe}
P.~Breitenlohner and D.~Maison, \emph{{Dimensional Renormalization of Massless
  Yang-Mills Theories}},  MPI-PAE-PTH-26-75, 1975.

\bibitem{Breitenlohner:1977hr}
P.~Breitenlohner and D.~Maison, \emph{{Dimensional Renormalization and the
  Action Principle}}, \href{https://doi.org/10.1007/BF01609069}{\emph{Commun.
  Math. Phys.} {\bfseries 52} (1977) 11}.

\bibitem{Breitenlohner:1975hg}
P.~Breitenlohner and D.~Maison, \emph{{Dimensionally Renormalized Green's
  Functions for Theories with Massless Particles. 1.}},
  \href{https://doi.org/10.1007/BF01609070}{\emph{Commun. Math. Phys.}
  {\bfseries 52} (1977) 39}.

\bibitem{Breitenlohner:1976te}
P.~Breitenlohner and D.~Maison, \emph{{Dimensionally Renormalized Green's
  Functions for Theories with Massless Particles. 2.}},
  \href{https://doi.org/10.1007/BF01609071}{\emph{Commun. Math. Phys.}
  {\bfseries 52} (1977) 55}.

\bibitem{Belusca-Maito:2020ala}
H.~Bélusca-Maïto, A.~Ilakovac, M.~Ma\dj{}or-Bo\v{z}inovi\'c and
  D.~Stöckinger, \emph{{Dimensional regularization and Breitenlohner-Maison/'t
  Hooft-Veltman scheme for $\gamma_5$ applied to chiral YM theories: full
  one-loop counterterm and RGE structure}},
  \href{https://doi.org/10.1007/JHEP08(2020)024}{\emph{JHEP} {\bfseries 08}
  (2020) 024}.

\bibitem{Martin:1999cc}
C.~P. Martin and D.~Sanchez-Ruiz, \emph{{Action principles, restoration of BRS
  symmetry and the renormalization group equation for chiral non-Abelian gauge
  theories in dimensional renormalization with a non-anticommuting
  $\gamma_5$}},
  \href{https://doi.org/10.1016/S0550-3213(99)00453-8}{\emph{Nucl. Phys.}
  {\bfseries B572} (2000) 387}.

\bibitem{Stockinger:2005gx}
D.~Stöckinger, \emph{{Regularization by dimensional reduction: consistency,
  quantum action principle, and supersymmetry}},
  \href{https://doi.org/10.1088/1126-6708/2005/03/076}{\emph{JHEP} {\bfseries
  03} (2005) 076}.

\bibitem{Hollik:2005nn}
W.~Hollik and D.~Stöckinger, \emph{{MSSM Higgs-boson mass predictions and
  two-loop non-supersymmetric counterterms}},
  \href{https://doi.org/10.1016/j.physletb.2006.01.030}{\emph{Phys. Lett.}
  {\bfseries B634} (2006) 63}.

\bibitem{Stockinger:2018oxe}
D.~Stöckinger and J.~Unger, \emph{{Three-loop MSSM Higgs-boson mass
  predictions and regularization by dimensional reduction}},
  \href{https://doi.org/10.1016/j.nuclphysb.2018.08.005}{\emph{Nucl. Phys.}
  {\bfseries B935} (2018) 1}
  [\href{https://arxiv.org/abs/1804.05619}{{\ttfamily 1804.05619}}].

\bibitem{Batalin:1977pb}
I.~A. Batalin and G.~A. Vilkovisky, \emph{{Relativistic S Matrix of Dynamical
  Systems with Boson and Fermion Constraints}},
  \href{https://doi.org/10.1016/0370-2693(77)90553-6}{\emph{Phys. Lett.}
  {\bfseries 69B} (1977) 309}.

\bibitem{Batalin:1981jr}
I.~A. Batalin and G.~A. Vilkovisky, \emph{{Gauge Algebra and Quantization}},
  \href{https://doi.org/10.1016/0370-2693(81)90205-7}{\emph{Phys. Lett.}
  {\bfseries 102B} (1981) 27}.

\bibitem{Batalin:1984jr}
I.~A. Batalin and G.~A. Vilkovisky, \emph{{Quantization of Gauge Theories with
  Linearly Dependent Generators}},
  \href{https://doi.org/10.1103/PhysRevD.28.2567,
  10.1103/PhysRevD.30.508}{\emph{Phys. Rev.} {\bfseries D28} (1983) 2567}.

\bibitem{Kraus:1997bi}
E.~Kraus, \emph{{Renormalization of the Electroweak Standard Model to All
  Orders}}, \href{https://doi.org/10.1006/aphy.1997.5746}{\emph{Annals Phys.}
  {\bfseries 262} (1998) 155}
  [\href{https://arxiv.org/abs/hep-th/9709154}{{\ttfamily hep-th/9709154}}].

\bibitem{Grassi:1999nb}
P.~A. Grassi, \emph{{Renormalization of nonsemisimple gauge models with the
  background field method}},
  \href{https://doi.org/10.1016/S0550-3213(99)00457-5}{\emph{Nucl. Phys. B}
  {\bfseries 560} (1999) 499}.

\bibitem{Hollik:2002mv}
W.~Hollik, E.~Kraus, M.~Roth, C.~Rupp, K.~Sibold and D.~Stöckinger,
  \emph{{Renormalization of the minimal supersymmetric standard model}},
  \href{https://doi.org/10.1016/S0550-3213(02)00538-2}{\emph{Nucl. Phys.}
  {\bfseries B639} (2002) 3}.

\bibitem{Piguet:1995er}
O.~Piguet and S.~P. Sorella, \emph{{Algebraic renormalization: Perturbative
  renormalization, symmetries and anomalies}},
  \href{https://doi.org/10.1007/978-3-540-49192-7}{\emph{Lect. Notes Phys.
  Monogr.} {\bfseries 28} (1995) 1}.

\bibitem{Becchi:1974md}
C.~Becchi, A.~Rouet and R.~Stora, \emph{{Renormalization of the Abelian
  Higgs-Kibble Model}}, \href{https://doi.org/10.1007/BF01614158}{\emph{Commun.
  Math. Phys.} {\bfseries 42} (1975) 127}.

\bibitem{Piguet:1980nr}
O.~Piguet and A.~Rouet, \emph{{Symmetries in Perturbative Quantum Field
  Theory}}, \href{https://doi.org/10.1016/0370-1573(81)90066-1}{\emph{Phys.
  Rept.} {\bfseries 76} (1981) 1}.

\bibitem{Kraus:1995jk}
E.~Kraus and K.~Sibold, \emph{{Rigid invariance as derived from BRS invariance:
  The Abelian Higgs model}}, \href{https://doi.org/10.1007/BF01566680}{\emph{Z.
  Phys. C} {\bfseries 68} (1995) 331}.

\bibitem{Haussling:1996rq}
R.~Haussling and E.~Kraus, \emph{{Gauge parameter dependence and gauge
  invariance in the Abelian Higgs model}},
  \href{https://doi.org/10.1007/s002880050521}{\emph{Z. Phys. C} {\bfseries 75}
  (1997) 739}.

\bibitem{Hahn:2000kx}
T.~Hahn, \emph{{Generating Feynman diagrams and amplitudes with FeynArts 3}},
  \href{https://doi.org/10.1016/S0010-4655(01)00290-9}{\emph{Comput. Phys.
  Commun.} {\bfseries 140} (2001) 418}.

\bibitem{Mertig:1990an}
R.~Mertig, M.~Bohm and A.~Denner, \emph{{FEYN CALC: Computer algebraic
  calculation of Feynman amplitudes}},
  \href{https://doi.org/10.1016/0010-4655(91)90130-D}{\emph{Comput. Phys.
  Commun.} {\bfseries 64} (1991) 345}.

\bibitem{Shtabovenko:2016sxi}
V.~Shtabovenko, R.~Mertig and F.~Orellana, \emph{{New Developments in FeynCalc
  9.0}}, \href{https://doi.org/10.1016/j.cpc.2016.06.008}{\emph{Comput. Phys.
  Commun.} {\bfseries 207} (2016) 432}.

\bibitem{Shtabovenko:2016whf}
V.~Shtabovenko, \emph{{FeynHelpers: Connecting FeynCalc to FIRE and
  Package-X}}, \href{https://doi.org/10.1016/j.cpc.2017.04.014}{\emph{Comput.
  Phys. Commun.} {\bfseries 218} (2017) 48}.

\bibitem{Patel:2016fam}
H.~H. Patel, \emph{{Package-X 2.0: A Mathematica package for the analytic
  calculation of one-loop integrals}},
  \href{https://doi.org/10.1016/j.cpc.2017.04.015}{\emph{Comput. Phys. Commun.}
  {\bfseries 218} (2017) 66}
  [\href{https://arxiv.org/abs/1612.00009}{{\ttfamily 1612.00009}}].

\bibitem{Machacek:1983tz}
M.~E. Machacek and M.~T. Vaughn, \emph{{Two Loop Renormalization Group
  Equations in a General Quantum Field Theory. 1. Wave Function
  Renormalization}},
  \href{https://doi.org/10.1016/0550-3213(83)90610-7}{\emph{Nucl. Phys.}
  {\bfseries B222} (1983) 83}.

\bibitem{Machacek:1983fi}
M.~E. Machacek and M.~T. Vaughn, \emph{{Two Loop Renormalization Group
  Equations in a General Quantum Field Theory. 2. Yukawa Couplings}},
  \href{https://doi.org/10.1016/0550-3213(84)90533-9}{\emph{Nucl. Phys.}
  {\bfseries B236} (1984) 221}.

\bibitem{Machacek:1984zw}
M.~E. Machacek and M.~T. Vaughn, \emph{{Two Loop Renormalization Group
  Equations in a General Quantum Field Theory. 3. Scalar Quartic Couplings}},
  \href{https://doi.org/10.1016/0550-3213(85)90040-9}{\emph{Nucl. Phys.}
  {\bfseries B249} (1985) 70}.

\bibitem{Zimmermann:1972te}
W.~Zimmermann, \emph{{Composite operators in the perturbation theory of
  renormalizable interactions}},
  \href{https://doi.org/10.1016/0003-4916(73)90429-6}{\emph{Annals Phys.}
  {\bfseries 77} (1973) 536}.

\bibitem{Zimmermann:1972tv}
W.~Zimmermann, \emph{{Normal products and the short distance expansion in the
  perturbation theory of renormalizable interactions}},
  \href{https://doi.org/10.1016/0003-4916(73)90430-2}{\emph{Annals Phys.}
  {\bfseries 77} (1973) 570}.

\bibitem{Lowenstein:1971vf}
J.~H. Lowenstein, \emph{{Normal product quantization of currents in Lagrangian
  field theory}}, \href{https://doi.org/10.1103/PhysRevD.4.2281}{\emph{Phys.
  Rev.} {\bfseries D4} (1971) 2281}.

\bibitem{Collins:1974da}
J.~C. Collins, \emph{{Normal Products in Dimensional Regularization}},
  \href{https://doi.org/10.1016/S0550-3213(75)80010-1}{\emph{Nucl. Phys.}
  {\bfseries B92} (1975) 477}.

\bibitem{Mertig:1998vk}
R.~Mertig and R.~Scharf, \emph{{TARCER: A Mathematica program for the reduction
  of two loop propagator integrals}},
  \href{https://doi.org/10.1016/S0010-4655(98)00042-3}{\emph{Comput. Phys.
  Commun.} {\bfseries 111} (1998) 265}
  [\href{https://arxiv.org/abs/hep-ph/9801383}{{\ttfamily hep-ph/9801383}}].

\bibitem{Misiak:1994zw}
M.~Misiak and M.~Munz, \emph{{Two loop mixing of dimension five flavor changing
  operators}}, \href{https://doi.org/10.1016/0370-2693(94)01553-O}{\emph{Phys.
  Lett. B} {\bfseries 344} (1995) 308}.

\bibitem{Chetyrkin:1997fm}
K.~G. Chetyrkin, M.~Misiak and M.~Munz, \emph{{Beta functions and anomalous
  dimensions up to three loops}},
  \href{https://doi.org/10.1016/S0550-3213(98)00122-9}{\emph{Nucl. Phys. B}
  {\bfseries 518} (1998) 473}.

\bibitem{Luthe:2017ttg}
T.~Luthe, A.~Maier, P.~Marquard and Y.~Schroder, \emph{{The five-loop Beta
  function for a general gauge group and anomalous dimensions beyond Feynman
  gauge}}, \href{https://doi.org/10.1007/JHEP10(2017)166}{\emph{JHEP}
  {\bfseries 10} (2017) 166}.

\bibitem{Schubert:1989wg}
C.~Schubert, \emph{{The Yukawa model as an example for dimensional
  renormalization with $\gamma_5$}},
  \href{https://doi.org/10.1016/0550-3213(89)90153-3}{\emph{Nucl. Phys. B}
  {\bfseries 323(2)} (1989) 478}.

\end{thebibliography}\endgroup

\end{document}